\documentclass[10pt,conference]{IEEEtran}
\ifCLASSOPTIONcompsoc
  \usepackage[nocompress]{cite}
\else
  \usepackage{cite}
\fi
\ifCLASSINFOpdf
  \usepackage[pdftex]{graphicx}
\else
\fi
\usepackage{listings}
\usepackage{xcolor}

\definecolor{codegreen}{rgb}{0,0.6,0}
\definecolor{codegray}{rgb}{0.5,0.5,0.5}
\definecolor{codepurple}{rgb}{0.58,0,0.82}
\definecolor{backcolour}{rgb}{0.95,0.95,0.92}

\lstdefinestyle{mystyle}{
    backgroundcolor=\color{backcolour},
    commentstyle=\color{codegreen},
    keywordstyle=\color{magenta},
    numberstyle=\tiny\color{codegray},
    stringstyle=\color{codepurple},
    basicstyle=\ttfamily\scriptsize,
    breakatwhitespace=false,         
    breaklines=true,                 
    captionpos=b,                    
    keepspaces=true,                 
    numbers=left,                    
    numbersep=5pt,                  
    showspaces=false,                
    showstringspaces=false,
    showtabs=false,    
    tabsize=2
}

\newcommand{\inlong}[1]{}

\usepackage{amsmath}

\graphicspath{ {figures/} }

\newcommand\pic[4]{
  \begin{figure}
  \centering
  \fbox{
    \includegraphics[width=#1\linewidth]{#2}
  }
  \caption{#3}
  \label{#4}
  \end{figure}
}

\newcommand\minipic[7]{
  \begin{figure}
  \centering
  
    \begin{minipage}{#1\linewidth}
      \centering
      \includegraphics[width=\linewidth]{#2}
      \subcaption{}
      \label{#7-a}
    \end{minipage}
    \hspace{#3}
    \begin{minipage}{#4\linewidth}
      \centering
      \includegraphics[width=\linewidth]{#5}
      \subcaption{}
      \label{#7-b}
    \end{minipage}
  
  \caption{#6}
  \label{#7}
  \end{figure}
}

\newcommand\minipicnosubcaption[7]{
  \begin{figure}
  \centering
  
    \begin{minipage}{#1\linewidth}
      \centering
      \includegraphics[width=\linewidth]{#2}
      \label{#7-a}
    \end{minipage}
    \hspace{#3}
    \begin{minipage}{#4\linewidth}
      \centering
      \includegraphics[width=\linewidth]{#5}
      \label{#7-b}
    \end{minipage}
  
  \caption{#6}
  \label{#7}
  \end{figure}
}

\newcommand\minipicthree[7]{
  \begin{figure*}
  \centering
  
    \begin{minipage}{#4\linewidth}
      \centering
      \includegraphics[width=\linewidth]{#1}
      \subcaption{}
      \label{#7-a}
    \end{minipage}
    \hspace{#5}
    \begin{minipage}{#4\linewidth}
      \centering
      \includegraphics[width=\linewidth]{#2}
      \subcaption{}
      \label{#7-b}
    \end{minipage}
    \hspace{#5}
    \begin{minipage}{#4\linewidth}
      \centering
      \includegraphics[width=\linewidth]{#3}
      \subcaption{}
      \label{#7-c}
    \end{minipage}
  
  \caption{#6}
  \label{#7}
  \end{figure*}
}

\newcommand\minipicfourbytwo[4]{%
    \def\figwidth{#1}%
    \def\gapwidth{#2}%
    \def\cpt{#3}%
    \def\mylabel{#4}%
    \minipicfourbytwocontinued
}

\newcommand\minipicfourbytwocontinued[8]{
  \begin{figure*}
  \centering
  
    \begin{minipage}{\figwidth\linewidth}
      \centering
      \includegraphics[width=\linewidth]{#1}
      \subcaption{}
      \label{\mylabel-a}
    \end{minipage}
    \hspace{\gapwidth}
    \begin{minipage}{\figwidth\linewidth}
      \centering
      \includegraphics[width=\linewidth]{#2}
      \subcaption{}
      \label{\mylabel-b}
    \end{minipage}
    \hspace{\gapwidth}
    \begin{minipage}{\figwidth\linewidth}
      \centering
      \includegraphics[width=\linewidth]{#3}
      \subcaption{}
      \label{\mylabel-c}
    \end{minipage}
    \hspace{\gapwidth}
    \begin{minipage}{\figwidth\linewidth}
      \centering
      \includegraphics[width=\linewidth]{#4}
      \subcaption{}
      \label{\mylabel-d}
    \end{minipage}
    
    \leavevmode\newline
  
    \begin{minipage}{\figwidth\linewidth}
      \centering
      \includegraphics[width=\linewidth]{#5}
      \subcaption{}
      \label{\mylabel-e}
    \end{minipage}
    \hspace{\gapwidth}
    \begin{minipage}{\figwidth\linewidth}
      \centering
      \includegraphics[width=\linewidth]{#6}
      \subcaption{}
      \label{\mylabel-f}
    \end{minipage}
    \hspace{\gapwidth}
    \begin{minipage}{\figwidth\linewidth}
      \centering
      \includegraphics[width=\linewidth]{#7}
      \subcaption{}
      \label{\mylabel-g}
    \end{minipage}
    \hspace{\gapwidth}
    \begin{minipage}{\figwidth\linewidth}
      \centering
      \includegraphics[width=\linewidth]{#8}
      \subcaption{}
      \label{\mylabel-h}
    \end{minipage}
  
  \caption{\cpt}
  \label{\mylabel}
  \end{figure*}
}

\usepackage{prettyref, url} 

\usepackage{booktabs}   
\usepackage{subcaption} 
                        
\usepackage{wrapfig}

\begin{document}
%

\title{Understanding Bounding Functions in Safety-Critical UAV Software}

\author{\IEEEauthorblockN{Xiaozhou Liang\IEEEauthorrefmark{1}, John Henry Burns\IEEEauthorrefmark{1}, Joseph Sanchez\IEEEauthorrefmark{1}, Karthik Dantu\IEEEauthorrefmark{2}, Lukasz Ziarek\IEEEauthorrefmark{2} and Yu David Liu\IEEEauthorrefmark{1}}\IEEEauthorblockA{\IEEEauthorrefmark{1}SUNY Binghamton, Binghamton, New York\\Email: {\tt \{xliang24,jburns11,jsanch49,davidL\}@binghamton.edu}}\IEEEauthorblockA{\IEEEauthorrefmark{2}SUNY Buffalo, Buffalo, New York\\Email: {\tt \{kdantu,lziarek\}@buffalo.edu}}
}


%


\maketitle

%
\IEEEpeerreviewmaketitle


\begin{abstract}

Unmanned Aerial Vehicles (UAVs) are an emerging computation platform known for their safety-critical need. In this paper, we conduct an empirical study on a widely used open-source UAV software framework, Paparazzi, with the goal of understanding the safety-critical concerns of UAV software from a bottom-up \emph{developer-in-the-field} perspective. We set our focus on the use of Bounding Functions (BFs), the runtime checks injected by Paparazzi developers on the range of variables. Through an in-depth analysis on BFs in the Paparazzi autopilot software, we found a large number of them (109 instances) are used to bound safety-critical variables essential to the cyber-physical nature of the UAV, such as its thrust, its speed, and its sensor values. The novel contributions of this study are two fold. First, we take a static approach to classify all BF instances, presenting a novel \emph{datatype-based} 5-category taxonomy with fine-grained insight on the role of BFs in ensuring the safety of UAV systems. Second, we dynamically evaluate the impact of the BF uses through a \emph{differential} approach, establishing the UAV behavioral difference with and without BFs. The two-pronged static and dynamic approach together illuminates a rarely studied design space of safety-critical UAV software systems.

\end{abstract}





\maketitle

\section{Introduction}
\label{section:introduction}

Unmanned aerial vehicles (UAVs) are an emerging platform with promising applications such as infrastructure inspection, precision agriculture, disaster search-and-rescue, and merchandise delivery.
Traditionally designed as a robotics and embedded system with minimal software support, the software stack of UAVs in recent years has been significantly enriched, making them a ``flying'' computer system in the genuine sense. Beyond the excitement, the main hurdle against the broader adoption of this promising technology is their stringent requirement on safety: any crash of the UAV is not only a computer safety problem, but also a public safety hazard.

Even though the safety-critical nature of UAVs is universally recognized, there is no universal definition of what safety really means for UAVs. Broadly, any behavior that deviates from  the ``intended behavior'' is a safety violation. Existing research~\cite{blanchet-pldi03,citenusmvforfcs,citenusmv,kloetzer2008fully,kress2009temporal,citecasetool,citer2u2}
generally takes a ``top-down'' approach: an expert may provide a specification of the intended behavior, either through domain knowledge, or through the wisdom from the broader domains of cyber-physical systems (CPS) or robotics. Once the specification is given -- whether in the form of invariants, constraints, pre-/post- conditions, or logic -- the safety of a UAV system can be verified, monitored, or enforced.

\subsection{An Empirical Perspective on UAV Software Safety}

In this paper, we take a bottom-up approach to empirically study the safety of UAV software. In a nutshell, we choose to listen to the UAV software developers \emph{in the field}, and reverse-engineer what they believe the most safety-critical software components are. Despite early UAV systems  often being developed in a proprietary fashion, recent trends in open-source development for UAV systems present an opportunity. For example, the software framework that serves as the focus of our empirical study, Paparazzi~\footnote{\url{https://wiki.paparazziuav.org/}}~\cite{r2u2-imav14}, is a popular open software (and hardware) ecosystem with more than a decade of development and numerous active contributors. It provides unified software support from autopilot to ground station, with diverse support for multi-copters, fixed-wing, helicopters and hybrid aircraft. If domain experts are the best source for understanding the ``intended behavior,'' what can we learn about UAV safety from UAV software developers themselves?

We focus on how \emph{Bounding Functions} (BF) are used in the Paparazzi autopilot software, arguably the most safety-critical components of the UAV software. A BF is a dynamic check inserted by programmers to ensure a variable -- which we call a \emph{Bounded Variable} (BV) -- stays within a prescribed range. For example, variable \texttt{gv\_z\_ref} in Paparazzi's navigation guidance module is frequently bounded by a BF within the range $[\texttt{cur\_z} - \texttt{GC\_MAX\_Z\_DIFF}, \texttt{cur\_z} + \texttt{GC\_MAX\_Z\_DIFF}]$. Here, the bounded variable \texttt{gv\_z\_ref} represents the altitude the UAV is guided to for the next time interval; variable \texttt{cur\_z} represents the current altitude of the UAV, and $\texttt{GC\_MAX\_Z\_DIFF}$ is a constant. Intuitively, this BF instance says that the UAV should not alter its altitude by $\texttt{GC\_MAX\_Z\_DIFF}$ or more within a time interval. This is aligned with our high-level understanding on UAV safety that an excessive change in altitude may jeopardize the stability of the UAV.

The premise of our approach is that the use of BFs is aligned with a UAV-specific safety concern. After all, the semantics of bounding a variable is akin to introducing an \emph{invariant} over the variable: the application of the bounding function is a no-op if the variable is already in the range, or an assignment to the variable with the bound value otherwise. If we take the programmer's perspective, the need to bound a variable is aligned with her concern that an ``out-of-range'' variable may cause errors in the program.

We take a \emph{two-pronged} approach in validating our premise. \emph{Statically}, we identify all BF instances in the source code, and provide a detailed \emph{datatype-based} taxonomy of the BF uses. We find a large number of BF uses indeed reflecting the safety-critical concerns of UAV systems (\S~\ref{sec:static}). \emph{Dynamically}, we perform a \emph{differential} simulation to illustrate the impact of BF uses on UAV behavior. We find BF uses, and their lack of use, do have impact on the dynamic trace of safety-critical values of the UAV, from trajectory to pose, and hence cyber-physical behavior of the UAV (\S~\ref{section:evaluation}). We now elaborate these two contributions in more detail.

\subsection{A Datatype-based Taxonomy}

A novelty of our empirical study is that we classify the use of BFs based on the \emph{datatype} of the BVs they intend to bound. In UAV software, a large number of values have primitive types such as \texttt{int} or \texttt{float}. A key insight gained in our exploration is that the BVs fall into a \emph{small} set of well-known UAV parameters reflecting their cyber-physical nature, which we call \emph{physical variables}. For example, we find a large number of \texttt{float}-type variables representing the 3 pose parameters that define the orientation of a UAV: the pitch, the roll, and the yaw. In other words, these variables carry higher-level semantics more than a floating point number. This insight recalls the classic programming abstraction of \emph{abstract data type} (ADT)~\cite{citeadt}: the \texttt{float} value above indeed logically encapsulates the floating number and a specification on what a pitch (or roll or yaw) parameter of a UAV should conform to. In this study, we classify our BF instances based on the logical \emph{datatypes} of their corresponding BVs, as follows:  

\begin{itemize}
    
 \item \textbf{Trajectory Management (TM)} BF instances that provide safe navigation to the UAV, mainly bounding physical variables such as \emph{position}, \emph{distance}, and \emph{heading}. 
            
    \item \textbf{Sensor Management (SM)} BF instances that provide valid sensor readings, bounding physical variables directly related to \emph{sensor values}. 

        \item \textbf{Speed and Acceleration Management (SAM)} BF instances that ensure safe \emph{speed} and \emph{acceleration} to engine, bounding these two physical variables. 
    
    \item \textbf{Engine Management (EM)} BF instances that provide safety to the engine by bounding  2 physical variables: the \emph{thrust} and \emph{throttle} of the engine.

    \item \textbf{Pose Management (PM)} BF instances that maintain safety for UAV orientation. These BF instances mainly bound 3 physical variables, \emph{pitch}, \emph{roll}, \emph{yaw} of the UAV.

\end{itemize}

Within each class, we perform an in-depth analysis on how BFs are used in Paparazzi, defined as \emph{use scenarios}. Taken the view of ADTs, each use scenario can be viewed as a \emph{specification} --- in the form of a BF --- of that datatype.
Overall, our novel datatype-based taxonomy can be summarized as ``not all floating point values (or integers) are created equal.'' By refining them into datatypes, their logical role in UAV software starts to emerge. As it turns out, except BF instances used for defining generic algorithms (such as control and geometry), the remaining BF instances \emph{all} fall into the 5 categories above. In other words, despite the large code base of UAV software and despite the numerous instances of BFs, UAV developers concentrate their efforts of performing dynamic checks on a small set of physical variables. This cannot be accidental: it is a conscious reminder that this small set of physical variables are likely to play a pivotal role in defining what being safety-critical means for UAV systems.

\subsection{A Differential Simulation}

To cross-validate whether our discovered BFs indeed have an impact on the correctness of UAV behavior, we perform a fine-grained simulation on the impact of BFs. We adopt a \emph{differential} approach: for each instance of BF use, we perform one simulation over the original Paparazzi program, and the other over the same program except that the BF is removed. At its core, our approach can be viewed as a form of A/B testing. The interesting design question lies in how difference is defined. Our approach relies on analyzing the difference over \emph{the traces of physical variables}, such as position traces (trajectories), pose traces, and speed traces. This approach, black-box in essence, is aligned with our intuition on the safety of UAV systems: if the UAV behaviors with the BF and without the BF are \emph{observably} different through the lens of physics, then the BF is likely impacting the safety of the UAV.

\subsection{Research Questions and Results}

In this paper, we report the first empirical study on the bounding function uses in UAV software. It complements existing top-down approaches with a bottom-up perspective focusing on answering two research questions:
\begin{itemize}
    
    \item \textbf{RQ1}: Can the BF instances be classified to logically reflect the use of safety-critical physical variables? 
    \item \textbf{RQ2}: Do BFs have impact on the dynamic cyber-physical behavior of UAV software? 
\end{itemize}

We identified 241 BF instances through analyzing Paparazzi's 2049 source files in autopilot software modules. We grouped 109 instances related to physical variables into the 5 categories (described earlier) most relevant to the safety of UAVs. Our dynamic differential analysis reveals that numerous BFs have observable impact on the trace of physical variables. More specifically, 30 out of 64 simulatable cases show difference in flight trajectory, pose, etc. This provides experimental justification for our BF-based approach: the use of BFs coincides with safety-critical behavior of UAVs. While conducting the trace-based analysis, we also uncovered a bug in Paparazzi, whose fix has been accepted.

Broadly and philosophically, our study is a quest for answers on what makes UAV software \emph{safety-critical}. The top-down approach taken by verification frameworks and tools defines safety as \emph{a priori} properties or invariants. To do so, one needs to resort to domain experts to come up with the definitions of these properties or invariants first. Our bottom-up developer-in-the-field approach identifies the use of BFs with a call for attention from developers, and the deviation in the dynamic traces of physical variables with a cause of safety concern ``as the developer's program says so.'' Overall, our approach and the existing approach complement each other: our approach discovers candidate invariants related to safety (but some may be deemed not by an “oracle” domain expert), whereas the existing approach focuses on invariants agreed upon \emph{a priori} (but they may be incomplete in the eyes of the “oracle” domain expert). The two approaches together converge on revealing the elusive essence of safety in UAV software.

Overall, this paper makes the following contributions:

\begin{itemize}
    \item the first ``developer-in-the-field'' empirical study on the safety-critical components of UAV software, based on bounding functions
    \item a datatype-based taxonomy on bounding function uses, focusing on physical variables
    \item a systematical differential analysis on the impact of BFs in UAV behavior through comparing and aligning traces of physical variables 
    \item a tool \texttt{PBF-Detector} (Paparazzi Bounding Function Detector) for automatically identifying BF instances in a real-world code base with complex compilation schemes (decentralized compilation with 78 makefiles mixed with pre-processing code generation) 
\end{itemize}
\section{A Primer on UAV Flight Control}
\label{section:background}

The most widely known UAVs fall into two categories: fixed-wing aircraft and rotary-wing aircraft. Fixed-wing aircraft are featured with special-shaped wings that can make use of forward airspeed to generate lift~\cite{citefixedwingdefinition}, while rotary-wing aircraft, also referred to as rotorcraft, use rotating wings called blades to fly~\cite{citerotarywingdefinition}. 

\subsection{Engine and Pose}

The driving force produced by the engine is commonly referred to as \emph{thrust} or \emph{throttle}. 
Engine management is directly associated with the \emph{speed} and the \emph{acceleration} of the UAV.
UAVs are rigid bodies operating in 3-D space. Therefore, their position can be represented by three numbers $(x, y, z)$ in a 3-D coordinate system. Similarly, their {\it pose} (orientation) is represented by three angles (also known as Euler angles) in the 3-D coordinate system. These angles are {\it roll, pitch and yaw}. The pose is also referred as the \emph{attitude}.
An illustration of the three angles can be found in Figure~\ref{fig:rollpitchyaw}. Fixed-wing aircraft vary their attitude by utilizing flight control surfaces. Rotorcraft vary the attitude by varying the rotational speeds of the motors spinning in opposite directions. 

\minipicnosubcaption{.55}{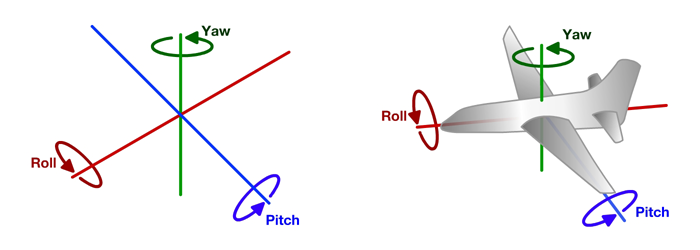}{10pt}{.25}{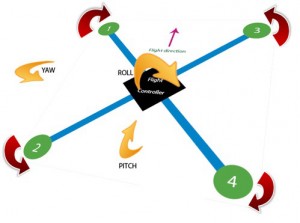}{A Visualization of UAV control (Left: attitude angles; Center: their application on a fixed-wing aircraft; Right: their application on a quadrotor)~\cite{prw-overview}}{fig:rollpitchyaw}

\subsection{Navigation}
\label{sec:navbk}

Navigating a UAV is usually split into two steps - path planning and trajectory planning. Path planning is the step of taking the objectives of a fight task. 
Path planning is usually application-dependent, written in the form of \emph{flight plans} in Paparazzi. 
For example, a typical flight plan may include a step-by-step description of take-off, a circle navigation task, and then landing. The flight plan is translated into a \emph{trajectory}.  The trajectory is defined through a series of \emph{waypoints}, positions in the 3-D space, with the Z-axis representing \emph{altitude}. Trajectory planning takes the next waypoint to be visited and plans a thrust and pose to set the UAV to reach that waypoint. Given the required thrust and pose, the flight controller controls the actuators (such as engines) to achieve that thrust and pose. While in motion, the UAV points to a direction, which is called \emph{heading}. A related concept is the \emph{course}, the direction that the UAV moves toward. Due to conditions such as wind, heading and course are not always the same. 

\subsection{Paparazzi Flight Controller Software}
Paparazzi UAV software suite is a collection of modules capable of flying on a variety of UAVs. It is highly configurable with various airframes, large suite of sensors, several controller algorithms as well as the ability to use the controller software in simulation and on real hardware.

\inlong{

\begin{figure}
\centering
\includegraphics[width=.5\linewidth]{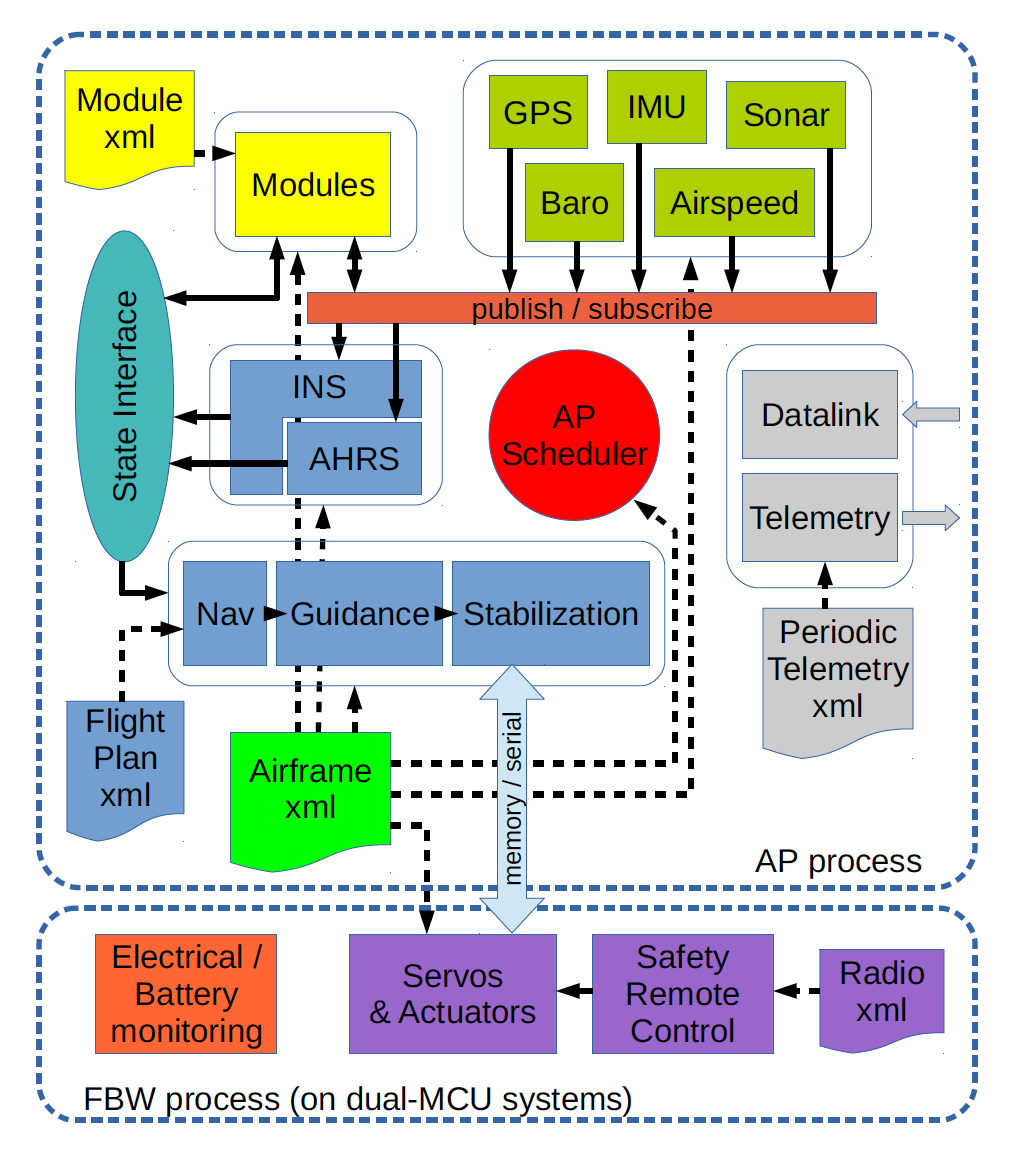}
\caption{Fixed-wing Airborne Software Architecture of Paparazzi Flight Controller as Shown in~\cite{r2u2-imav14}}
\label{fig:parch}
\end{figure}

We will briefly repeat the description of the individual modules in the software from~\cite{r2u2-imav14} for clarity as shown in Fig.~\ref{fig:parch}. 
}
The autopilot software is capable of integrating with several sensors, such as GPS, Inertial Measurement Unit (IMU), Sonar, and barometer. 
Sensor values are fed into the Inertial Navigation System (INS) that estimates position, speed, and acceleration of the UAV. 
Similarly, the Attitude and Heading Reference System (AHRS) performs attitude estimation.
Together, the INS and the AHRS help the flight controller keep an estimate of the state of the UAV. This state is then used to control the UAV through the guidance and stabilization modules.

As is the case for all aerodynamic systems, control-theoretic algorithms are widely used to provide feedback control in UAVs' stability management and autonomous control. Two popular algorithms used by Paparazzi are Proportional Integral Derivative (PID) control \cite{citepid} and Incremental Nonlinear Dynamic Inversion (INDI) \cite{citeindi}.

\section{Understanding Bounding Functions Statically} 
\label{sec:static}

In this section, we describe our effort in understanding BF uses in Paparazzi through a detailed analysis on the source code, providing answers to \textbf{RQ1}. The centerpiece of this study is a \emph{taxonomy} that classifies BF uses based on the physical variables they are applied to, in \S~\ref{section:BF_Classification}. Before we detail this result, we start with a description of our taxonomy rationale in \S~\ref{sec:prin}, and methodology in \S~\ref{sec:met}. 

\subsection{The Rationale of Classification}
\label{sec:prin}

UAVs are cyber-physical systems that interact with the physical world. Their safety is defined with respect to this interaction, i.e., their behavior in the physical world. Our classification of BFs is based on this observation and thus derived from the \emph{datatypes of physical variables} associated with the BFs. Our five-category taxonomy corresponds to the main functionalities of the UAV that define or impact interaction with the physical world. By organizing BF uses in this manner, we believe that this study will be useful for future UAV control software as they will still need to fundamentally interact with the physical world in the same manner: they will need to navigate (trajectory), control their navigation (speed and acceleration), understand their surroundings (sensors), understand their orientation with respect to their surroundings (pose), and manage their locomotion (motors). 
As our study shows, the vast majority of BFs in Paparazzi revolve around these five types of physical variables. This cannot be accidental: these five types of cyber-physical interactions are essential to the nature of UAV software.

\subsection{Methodology}
\label{sec:met}

\begin{table}
  \small
  \caption{Bounding Functions in Paparazzi}
  \label{tab:boundingfunctionclassification}
  \begin{tabular}{p{3cm}p{5cm}}
    \toprule 
    Bound forms & Function names \\
    \midrule
    double-ended bounds & Bound \newline BoundInverted \newline BoundWrapped \newline VECT3{\_}BOUND{\_}CUBE \newline VECT3{\_}BOUND{\_}BOX \newline EULERS{\_}BOUND{\_}CUBE \newline RATES{\_}BOUND{\_}CUBE \newline RATES{\_}BOUND{\_}BOX \newline Clip \\
    absolute bounds & BoundAbs \newline RATES{\_}BOUND{\_}BOX{\_}ABS \newline DeadBand \newline ClipAbs \\
    upper bounds & BoundUpper \\
    normalization & FLOAT{\_}ANGLE{\_}NORMALIZE \newline INT32{\_}ANGLE{\_}NORMALIZE \newline INT32{\_}COURSE{\_}NORMALIZE \newline NormRadAngle \\
    special bounds & SATURATE{\_}SPEED{\_}TRIM{\_}ACCEL \\
    \bottomrule
\end{tabular}
\end{table}

\paragraph{BF Identification}

We have developed a compiler pass, implemented as a Clang plugin~\footnote{\url{https://clang.llvm.org/}}, to identify BF instances in the Paparazzi code base. 
Our plugin defines a baseline framework, \texttt{PBF-Detector}, for future research with advanced program analysis and optimization. Our analysis focuses on Paparazzi's autopilot software modules, in the \texttt{sw/airborne} directory, version v5.14.0\_stable. 

For our goal of identifying BFs, Paparazzi presents a unique advantage: a set of pre-defined BFs in the forms of C macros  consistently used by Paparazzi developers. Our study focuses on the use of these macros, 19 in total as listed in Table~\ref{tab:boundingfunctionclassification}. These macros are manually identified by inspecting all \texttt{.h} files, and a macro qualifies if it bounds a variable within a given range. Some BFs are general, such as \texttt{Bound}, while others are more specific.
Our Clang plugin parses C files to identify the 19 forms of BFs in the AST. One technical hurdle is that macros are expanded in Clang before the AST is generated. To address this, we have redefined 19 corresponding C functions to the macros in Table~\ref{tab:boundingfunctionclassification}. The Paparazzi source remains unchanged with a small number of exceptions that we documented at our project website (see URL in \S~\ref{section:finaldiscussion}).

\paragraph{Makefile-Aware Identification}

A significant engineering challenge in analyzing Paparazzi's code base results from the complex compilation process inherent in Paparazzi. Unlike high-level applications where the compilation process is often a ``one-off'' process that reaches all files in all folders, embedded system software like Paparazzi must consider diverse configurations with complex customization and platform-dependent cross-compilation. Paparazzi adopts a hierarchical compilation with 78 C Makefiles distributed at various levels of the Paparazzi directories, and the dependencies between Makefile targets are complex. To further complicate the matter, many programs are generated on the fly during the compilation process with generators written in OCaml and Python. 

Our compiler pass is \emph{Makefile-aware}: we modified the decentralized Makefiles, and as a result, \texttt{PBF-Detector} can faithfully follow the same dependencies as in compilation. This not only allows us to reach all source code that can be reached by Paparazzi compilation, but also reach it in a semantic-aware manner: every name on the AST of every reachable file must have been defined (because the program compiles!). The \texttt{PBF-Detector} modification to handle hierarchical Makefiles in our compiler analysis was labor-intensive, but it is rewarding for building a toolchain for Paparazzi to integrate with Clang/LLVM.

\paragraph{BF Selection}

In total, we identified 241 instances of BFs from autopilot program modules spanning 2049 files in 331K LOC of Paparazzi source code. We further cross-validated the number of instances through a text-based search. Among them, our study excludes instances not directly related to the safety of UAV software, which fall into 3 categories: (a) 71 BF instances in core control algorithms (PID/INDI). These BF instances are part of the algorithm design, such as PID and INDI; they are ``generic'' in nature and do not vary from a UAV implementation to a non-UAV implementation. As a standard robotics problem, bounding and tuning generic control parameters is an independent and well-studied problem~\cite{aastrom1995pid}. It should be made clear that we only leave out generic control algorithm BF uses here: if a physical variable, say the roll value of the UAV, relies on the PID control and is bounded while interacting with the PID, it \emph{is} included in our study. (b) 45 BF instances used for geometric transformation. These BFs occur as parts of the trigonometry-based algorithms solely related to geometry.  For example, a common use is to normalize an angle within the range of (-2$\pi$ , 2$\pi$). (c) 15 BF instances in vision/image processing algorithms and 1 instance used for the remote control switch. For instance, BFs frequently occur for managing auto white balance, image refinement, sub-pixel resolution, and auto exposure. For any vision BF instances that impact control algorithms (e.g.,  optic-flow-based landing), we have included them into our study. 
Overall, our guiding principle here is to conservatively leave out BF instances unrelated to the safety-critical nature of UAV software, and when in doubt, an instance is included in our study. 
We have documented every BF instance for cross reference, including those we left out in the study on our project website.

There are two take-away messages from our BF selection process. On one hand, it shows some BF instances are not aligned with our intuition of safety-critical concerns. In that sense, these instances are the ``false positives'' to the premise of our empirical study. On the other hand, the more striking observation is that once the well-carved categories of (a)(b)(c) BF instances are removed, \emph{every} remaining instance fits nicely with one of 5 categories intimately linked to the safety of UAVs, as we shall see next.

\subsection{A Taxonomy of Bounding Function Uses}
\label{section:BF_Classification}

For the remaining 109 BF instances, we conducted an in-depth manual inspection, understanding the functionality of the program fragment each appears, and the purpose of each BF. As it turns out, all fit nicely into the 5-category taxonomy, which we present in Table~\ref{tab:boundingfunctionusesclassification}. In this table, observe that we further refine each category into a number of \emph{use scenarios}. If each category is intuitively viewed as an ADT, each use scenario serves as a specified behavior of that ADT. In the rest of this section, we focus on trajectory management and sensor management as examples to demonstrate our approach. A description of all categories with the same level of detail can be found in a technical report at our project website. 

\begin{table}
  \small
  \caption{Classification of Bounding Function Uses}
  \label{tab:boundingfunctionusesclassification}
  \begin{tabular}{p{0.8cm}|p{3cm}|p{2.3cm}|p{0.9cm}}
  \hline
Cate-\newline gory & Use Scenario & Datatype & Occur-\newline rence \\ \hline
{TM}             & Safe Leg Distance in \newline Guidance               & Distance (X and Y Axis)               & 4          \\ \cline{2-4} 
                                                   & Safe Heading Change                       & Heading Change                        & 3          \\ \cline{2-4} 
                                                   & Safe Homing                               & Distance (X and Y Axis)               & 3          \\ \cline{2-4} 
                                                   & Safe Altitude Change                      & Distance (Z Axis)                     & 1          \\ \hline
{SM}               & Safe Sensor Fusion                          & Weight for Sensor Fusion              & 6          \\ \cline{2-4} 
                                                   & Safe Sensor Reading \newline Interval              & Time Interval                         & 1          \\ \cline{2-4} 
                                                   & Safe   Sensor Readings                      & Sensor Reading                        & 1          \\ \hline
{SAM} & Safe Acceleration Request as Engine Input & Acceleration                          & 8          \\ \cline{2-4} 
                                                   & Safe Acceleration for \newline Navigation          & Acceleration                          & 4          \\ \cline{2-4} 
                                                   & Safe Remote User \newline Speed Input              & Speed                                 & 3          \\ \cline{2-4} 
                                                   & Safe   Wind Speed                           & Speed                                 & 2          \\ \hline
{EM}               & Safe Motor Mixing                           & Thrust/Throttle                       & 8          \\ \cline{2-4} 
                                                   & Safe   Landing                              & Thrust/Throttle                       & 7          \\ \cline{2-4} 
                                                   & Safe Motor Speed \newline Change                   & RPM                                   & 5          \\ \cline{2-4} 
                                                   & Collision Avoidance                       & Thrust/Throttle                       & 1          \\ \hline
{PM}                 & Safe Pose Change \newline Rate                       & Pitch/Roll/Yaw                        & 37         \\ \cline{2-4} 
                                                   & Safe Pose Maintenance                     & Pitch/Roll/Yaw                        & 12         \\ \cline{2-4} 
                                                   & Safe Pose Change \newline Time Interval            & Pitch/Roll/Yaw \newline Change Time \newline Interval & 2          \\ \cline{2-4} 
                                                   & Safe   Turn Coordination                    & Roll                                  & 1          \\ \hline
\end{tabular}
\end{table}

\inlong{
\subsection{Engine Management}
\label{section:motormanagement}

The heart of UAVs are the engines. For motor-based engines, the force it provides is correlated with its number of \emph{Revolutions Per Minute} (RPM). UAV systems typically quantify the force through the \emph{throttle} and the \emph{thrust}. Strictly speaking, the throttle is an engine component that manipulate the thrust output (i.e., the force), but the two are often used interchangeably in the context of UAVs. In Paparazzi, throttle/thrust is represented by a value ranging between [0, 1], with 1 aligned with our intuition of ``full force.'' We identified 21 instances of BFs applied for engine management, over the physical variables of thrust/throttle and RPM. They are divided into the following use scenarios.

\subsubsection{Safe Landing}

\underline{(a) Use Context}:  
A "smooth" landing of the UAV coincides with the exponential reduction of both height and vertical velocity. According to optic flow theory \cite{citeflowdivergence}, this can be achieved through maintaining a constant \emph{flow divergence}, which can be effectively managed by thrust control. 
\underline{(b) Datatype}: thrust/throttle. 
\underline{(c) The Need for BFs}: as landing generally requires a reduction on the thrust, the upper bound is easily understood. The lower bound, however, is equally important: an excessively low thrust may cause stall, a catastrophic event that UAVs do not have sufficient lift to counter gravity.  
\underline{(d) Example}: 
In this code snippet~\footnote{Function \texttt{PID\_divergence\_control} at \url{https://github.com/paparazzi/paparazzi/blob/master/sw/airborne/modules/ctrl/optical_flow_landing.c}}, the thrust value output by the PID controller for optic flow is bounded within the range of \begin{center}$[$\texttt{0.25 * of{\_}landing{\_}ctrl.nominal{\_}thrust * MAX{\_}PPRZ}$,$ \texttt{MAX{\_}PPRZ}$]$\end{center} \noindent where \texttt{of{\_}landing{\_}ctrl.nominal{\_}thrust} is the nominal thrust around which the PID control operates, and $\texttt{MAX{\_}PPRZ}$ is a constant that can be customized by Paparazzi users. 
\underline{(e) Occurrence}: 7 instances.

\subsubsection{Safe Motor Mixing}

\underline{(a) Use Context}:  
For rotorcraft with at least 2 rotors, the throttle of each motor is managed by its controller, and ``mixed'' together according to the layout among the motors~\cite{citemotormixing} to compute the overall throttle of the UAV.
\underline{(b) Datatype}: thrust/throttle. 
\underline{(c) The Need for BFs}:
Unlike single motors where the change of controller output can be managed by the PID/INDI-like control algorithms themselves, the output of motor mixing may experience rapid change. Drastic change of UAV thrust/throttle may cause instability for the maneuver of UAVs.  
\underline{(d) Example}: In this code snippet~\footnote{Function \texttt{motor\_mixing\_run} at \url{https://github.com/paparazzi/paparazzi/blob/master/sw/airborne/subsystems/actuators/motor_mixing.c}}, the change in throttle by consecutive motor mixing outputs is represented by a variable named \texttt{saturation{\_}offset}, which is in turn bound to a range. \underline{(e) Occurrence}: 8 instances.

\subsubsection{Safe Motor Speed Change}

\underline{(a) Use Context}: Motor speed is correlated with the thrust/throttle it may produce. 
\underline{(b) Datatype}: RPM. 
\underline{(c) The Need for BFs}: A drastic change in motor speed may not only cause a drastic change in thrust (which will cause instability for UAVs as we described earlier), but also may cause damage to the motor itself. In manually controlled UAVs, users may adjust the throttle curve settings, a discrete representation of different levels of throttle. The change in motor speed must be bounded to avoid excessive user-inputed change.
\underline{(d) Example}:  In this code snippet~\footnote{Function \texttt{throttle\_curve\_run} at \url{https://github.com/paparazzi/paparazzi/blob/master/sw/airborne/modules/helicopter/throttle_curve.c}}, the change in RPM is represented by a variable named \texttt{rpm{\_}diff}, which is in turn bound to the range of 
\begin{center}$[$\texttt{-THROTTLE{\_}CURVE{\_}RPM{\_}INC{\_}LIMIT} / 512$,$ \texttt{THROTTLE{\_}CURVE{\_}RPM{\_}INC{\_}LIMIT} / 512$]$,\end{center} where \texttt{THROTTLE{\_}CURVE{\_}RPM{\_}INC{\_}LIMIT} is a customizable limit for RPM increment.
\underline{(e) Occurrence}: 5 instances.

\subsubsection{Collision Avoidance}

\underline{(a) Use Context}: Obstacle and collision avoidance is an essential safety-critical concern of UAVs. In Paparazzi, UAV obstacle avoidance is based on potential field \cite{citeobstacleavoidance}. 
A throttle control using potential field is utilized to compute the desired throttle/thrust to avert aircraft collision.  
\underline{(b) Datatype}: throttle/thrust.
\underline{(c) The Need for BFs}: A high throttle level would lead to a overly large speed of the UAV. In this case, a collision may occur because of the inertia of the UAV even though potential field is used for avoidance.
\underline{(d) Example}: In Listing \ref{code:potentialc}, the desired throttle level \texttt{cruise} computed using potential field is bounded.
Note that even though the variables here may indicate \texttt{cruise} is related to cruise speed, the semantics beneath the variable is indeed a throttle value. For example, constants $\texttt{V{\_}CTL{\_}AUTO{\_}THROTTLE{\_}MIN{\_}CRUISE{\_}THROTTLE}$ and $ \texttt{V{\_}CTL{\_}AUTO{\_}THROTTLE{\_}MAX{\_}CRUISE{\_}THROTTLE}$ in the default implementation of Paparazzi are set at 0.2 and 1.0, respectively.
\underline{(e) Occurrence}: 1 instance.

\begin{lstlisting}[language=C, caption={An Example on Throttle/Thrust Bounding for Collision Avoidance\protect\footnotemark}, captionpos=b, label={code:potentialc}]
int potential_task(void)
{
  ...
  
  // speed loop
  float cruise = V_CTL_AUTO_THROTTLE_NOMINAL_CRUISE_THROTTLE;
  cruise += -force_speed_gain * (potential_force.north * ch + potential_force.east * sh);
  Bound(cruise, V_CTL_AUTO_THROTTLE_MIN_CRUISE_THROTTLE, V_CTL_AUTO_THROTTLE_MAX_CRUISE_THROTTLE);
  potential_force.speed = cruise;
  v_ctl_auto_throttle_cruise_throttle = cruise;
  
  ...
}
\end{lstlisting}
\footnotetext{Function \texttt{potential\_task} at \url{https://github.com/paparazzi/paparazzi/blob/master/sw/airborne/modules/multi/potential.c}}

\subsection{Speed and Acceleration Management}
\label{section:speedandaccel}

The force generated by the engine directly impact the \emph{acceleration} of the UAVs, which in turn impacts their \emph{speed}. We identified 17 instances of BFs applied for speed and acceleration management. They are divided into 4 use scenarios.

\subsubsection {Safe Acceleration for Navigation}

\underline{(a) Use Context}:  A positive acceleration is needed for speeding up the UAV, whereas a negative one manages to slow it down. Acceleration is essential for the take-off and landing of UAVs.
\underline{(b) Datatype}: acceleration. 
\underline{(c) The Need for BFs}: an extremely high value of the acceleration may negatively impact the stability of UAVs.
\underline{(d) Example}: In this code snippet~\footnote{Function \texttt{gh\_saturate\_ref\_accel} at \url{https://github.com/paparazzi/paparazzi/blob/master/sw/airborne/firmwares/rotorcraft/guidance/guidance_h_ref.c}}, horizontal acceleration in the two-dimensional reference model, \texttt{gh{\_}ref.accel}, is bounded on the X and Y directions respectively. The bounds are set by another variable, \texttt{gh{\_}ref.max{\_}accel}. This is an example where the bound is not a constant but another dynamic program value, i.e., a \emph{dynamic bound}. \underline{(e) Occurrence}: 4 instances.

\subsubsection {Safe Acceleration Request as Engine Input}

\underline{(a) Use Context}: As acceleration must be realized by engine throttle/thrust changes, different navigation program modules may request a desirable acceleration to the control algorithm of the engine itself. 
\underline{(b) Datatype}: acceleration. 
\underline{(c) The Need for BFs}: an excessive input may require the engine to produce a large throttle/thrust, causing a flame-out in the engine. 
\underline{(d) Example}: In this code snippet~\footnote{Function \texttt{v\_ctl\_climb\_loop} at \url{https://github.com/paparazzi/paparazzi/blob/master/sw/airborne/firmwares/fixedwing/guidance/energy_ctrl.c}}, the target vertical acceleration is represented by variable \texttt{v{\_}ctl{\_}desired{\_}acceleration}, and it is bounded within  $[$\texttt{-v{\_}ctl{\_}max{\_}acceleration}$,$ \texttt{v{\_}ctl{\_}max{\_}acceleration}$]$, where \texttt{v{\_}ctl{\_}max{\_}acceleration} is the dynamic upper bound for a UAV's vertical acceleration.
\underline{(e) Occurrence}: 8 instances.

\subsubsection{Safe Wind Speed}

\underline{(a) Use Context}: While operating in the physical environments, UAVs need to consider the wind speed while calculating its ground speed, an important parameter that impacts its navigation, take-off, and landing.   
\underline{(b) Datatype}: speed. 
\underline{(c) The Need for BFs}: Wind can be ``gusty'' in nature, so that the measurement received from sensors may be unreliable. If the measured wind speed is excessive, it may severely impact how navigation trajectory is computed, and worse, jeopardize the safety of take-off and landing. 
\underline{(d) Example}: Listing \ref{code:navglsc} shows a code snippet for UAV landing. Here the ground speed is calculated with wind speed in consideration. The wind speed \texttt{wind{\_}on{\_}final} is bounded within the range $[$\texttt{-MAX{\_}WIND{\_}ON{\_}FINAL}$,$ \texttt{MAX{\_}WIND{\_}ON{\_}FINAL}$]$ at line 6. If bounding was not performed, the ground speed can be excessive, impacting the decision making process during landing. 
\underline{(e) Occurrence}: 2 instances.

\begin{lstlisting}[language=C, caption=An Example on Wind Speed Bounding~\protect\footnotemark, captionpos=b, label={code:navglsc}]
static inline bool gls_compute_TOD(uint8_t _af, uint8_t _sd, uint8_t _tod, uint8_t _td)
{
  ...
  
  float wind_on_final = wind_norm * (((td_af_x * wind->y) / (td_af * wind_norm)) + ((td_af_y * wind->x) / (td_af * wind_norm)));
  Bound(wind_on_final, -MAX_WIND_ON_FINAL, MAX_WIND_ON_FINAL);
  gs_on_final = target_speed - wind_on_final;
  
  ...
}
\end{lstlisting}
\footnotetext{Function \texttt{gls\_compute\_TOD} at \url{https://github.com/paparazzi/paparazzi/blob/master/sw/airborne/modules/nav/nav_gls.c}}

\subsubsection{Safe Remote User Speed Input}

\underline{(a) Use Context}:  For public safety reasons, UAVs with an autopilot board may often be required to operate with operator-guided remote control through governmental regulations.   
\underline{(b) Datatype}: speed. 
\underline{(c) The Need for BFs}: The operator may manually input a speed change that may impact the stability of a UAV.
\underline{(d) Example}: In this code snippet~\footnote{Function \texttt{guidance\_v\_read\_rc} at \url{https://github.com/paparazzi/paparazzi/blob/master/sw/airborne/firmwares/rotorcraft/guidance/guidance_v.c}}, a variable \texttt{guidance{\_}v{\_}rc{\_}zd{\_}sp} represents the vertical speed change inputted by the operator from remote control. It is bounded using a \emph{deadband}, i.e., the safety threshold, as 
[$ \texttt{-GUIDANCE{\_}V{\_}CLIMB{\_}RC{\_}DEADBAND}$,$ \texttt{GUIDANCE{\_}V{\_}CLIMB{\_}RC{\_}DEADBAND}$]. 
\underline{(e) Occurrence}: 3 instances

\subsection{Pose Management}
\label{section:rotationcommandmanagement}

Maintaining and maneuvering the pose of UAVs is essential for supporting key behaviors of UAVs, from take-off, to hover, to turn, to landing. The physical variables that represent the pose are \emph{pitch}, \emph{roll}, and \emph{yaw}. When the change of the pose matters, pose management also considers the \emph{rate} or \emph{interval} of change for the three physical variables. We identified 52 instances of BFs applied for pose management, divided into 4 use scenarios.

\subsubsection{Safe Pose Maintenance}

\underline{(a) Use Context}: Maintaining a pose plays a central role in UAV flight tasks such as follows. First, the take-off and landing requires a positive/negative value on pitch. Second, hover control, i.e., maintaining the position of a UAV in the 3-D space, is achieved through continuous correction on the pose through controllers. Third, obstacle avoidance also requires a change in pose, so that the trajectory may change as a result. 
\underline{(b) Datatype}: pitch/roll/yaw. 
\underline{(c) The Need for BFs}: Not all poses are safe for UAVs from an aerodynamic perspective. For instance, excessive pitch may cause stall. As another example, excessive roll may cause the UAV to flip.  
\underline{(d) Example}: In this code snippet~\footnote{Function \texttt{OA\_update} at \url{https://github.com/paparazzi/paparazzi/blob/master/sw/airborne/modules/obstacle_avoidance/guidance_OA.c}}, the roll and pitch outputs of the obstacle avoidance algorithm are bounded by $[$\texttt{-CMD{\_}OF{\_}SAT}$,$ \texttt{CMD{\_}OF{\_}SAT}$]$, where \texttt{CMD{\_}OF{\_}SAT} is the maximum angle.    \underline{(e) Occurrence}: 12 instances.

\subsubsection{Safe Turn Coordination}
\label{subsection:turncoordination}

\underline{(a) Use Context}: a concrete instance of safe pose maintenance, but with a unique constraint, is UAV turn coordination. When a UAV makes a turn, its roll value must be adjusted so that the UAV ``banks'' to an angle. 
\underline{(b) Datatype}: roll. 
\underline{(c) The Need for BFs}: a unique problem with turning is \emph{side-slipping}: if a roll angle is excessive, the UAV may move sideways rather than a smooth turn. 
\underline{(d) Example}: In this code snippet~\footnote{Function \texttt{gvf\_control\_2D} at \url{https://github.com/paparazzi/paparazzi/blob/master/sw/airborne/modules/guidance/gvf/gvf.c}}, the target roll angle is bounded within the range of $[$\texttt{-h{\_}ctl{\_}roll{\_}max{\_}setpoint}$,$ \texttt{h{\_}ctl{\_}roll{\_}max{\_}setpoint}$]$. \underline{(e) Occurrence}: 1 instance.

\subsubsection{Safe Pose Change Rate}
\label{sub:posechangerate}

\underline{(a) Use Context}:  In UAV pose management, an attitude reference model is used to estimate the pose of the UAV and calculate the desired angular rate of change. In Paparazzi, INDI is commonly used for this purpose. 
\underline{(b) Datatype}: pitch/roll/yaw change rate. 
\underline{(c) The Need for BFs}: a drastic change in pose may impact the stability of the UAV.
\underline{(d) Example}: In the code snippet in Listing \ref{code:stabilizationindisimplec}, the desired angular rate on yaw direction \texttt{rate{\_}ref{\_}r} is bounded into the range \begin{center}$[$\texttt{-indi.attitude{\_}max{\_}yaw{\_}rate}$,$ \texttt{indi.attitude{\_}max{\_}yaw{\_}rate}$]$,\end{center} where \texttt{indi.attitude{\_}max{\_}yaw{\_}rate} is the maximum yaw rate in pose control.
\underline{(e) Occurrence}: 37 instances.

\begin{lstlisting}[language=C, caption=An Example on Safe Pose Change\protect\footnotemark, captionpos=b, label={code:stabilizationindisimplec}]
static inline void stabilization_indi_calc_cmd(int32_t indi_commands[], struct Int32Quat *att_err, bool rate_control)
{
  ...
	
  //This separates the P and D controller and lets you impose a maximum yaw rate.
  float rate_ref_r = indi.reference_acceleration.err_r * QUAT1_FLOAT_OF_BFP(att_err->qz) / indi.reference_acceleration.rate_r;
  BoundAbs(rate_ref_r, indi.attitude_max_yaw_rate);
  indi.angular_accel_ref.r = indi.reference_acceleration.rate_r * (rate_ref_r - rates_for_feedback.r);
  
  ...
}
\end{lstlisting}
\footnotetext{Function \texttt{stabilization\_indi\_calc\_cmd} at \url{https://github.com/paparazzi/paparazzi/blob/master/sw/airborne/firmwares/rotorcraft/stabilization/stabilization_indi_simple.c}}

\subsubsection{Safe Pose Change Time Interval}

\underline{(a) Use Context}: A variant of the use scenario described in \S~\ref{sub:posechangerate} is that pose change may be quantified by the time interval between two adjustments. 
\underline{(b) Datatype}: pitch/roll/yaw change time interval. 
\underline{(c) The Need for BFs}: same as the use scenario described in \S~\ref{sub:posechangerate}. 
\underline{(d) Example}: In Listing \ref{code:stabilizationattitudercsetpointc}, for instance, \texttt{dt} represents the time interval since last call of the current function. Before used to compute the yaw angle \texttt{sp->psi} at line 12, it is bounded within 0.5 seconds to ensure a moderate increment of the target angle.
\underline{(e) Occurrence}: 2 instances.

\begin{lstlisting}[language=C, caption=An Example on Safe Pose Change Time Interval~\protect\footnotemark, captionpos=b, label={code:stabilizationattitudercsetpointc}]
void stabilization_attitude_read_rc_setpoint_eulers(struct Int32Eulers *sp, bool in_flight, bool in_carefree, bool coordinated_turn)
{
  ...
  
  /* calculate dt for yaw integration */
  float dt = get_sys_time_float() - last_ts;
  /* make sure nothing drastically weird happens, bound dt to 0.5sec */
  Bound(dt, 0, 0.5);

  /* do not advance yaw setpoint if within a small deadband around stick center or if throttle is zero */
  if (YAW_DEADBAND_EXCEEDED() && !THROTTLE_STICK_DOWN()) {
    sp->psi += get_rc_yaw() * dt;
    INT32_ANGLE_NORMALIZE(sp->psi);
  }
  
  ...
}
\end{lstlisting}
\footnotetext{Function \texttt{stabilization\_attitude\_read\_rc\_setpoint\_eulers} at \url{https://github.com/paparazzi/paparazzi/blob/master/sw/airborne/firmwares/rotorcraft/stabilization/stabilization_attitude_rc_setpoint.c}}

}
\subsubsection{Trajectory Management}

To follow a trajectory, the UAV needs to follow waypoints, including turning occasionally (in the horizontal direction) and changing altitude (in the vertical direction). The physical variables related to trajectory management are \emph{distance} and \emph{heading} (change). We identified 11 instances of BFs applied for trajectory management, which we divided into 4 use scenarios.

\paragraph{Safe Homing}

\underline{(a) Use Context}:  
After performing the flight task, the UAV should go back to the ground station.
\underline{(b) Datatype}: distance (X and Y axis).
\underline{(c) The Need for BFs}: to avoid catastrophic consequences due to battery drain, a UAV (generally) should not fly too far way from the ground station. 
\underline{(d) Example}: In this code snippet~\cite{citefunctionnavmovewaypoint}, 
the distance between the UAV waypoint and the home waypoint is computed, and bounded by variable \texttt{max{\_}dist{\_}from{\_}home}, the maximum distance between them. \underline{(e) Occurrence}: 3 instances.

\paragraph{Safe Altitude Change}

\underline{(a) Use Context}:  UAV systems fly in a 3-D space; altitude change is a basic task.
\underline{(b) Datatype}: distance (Z axis).
\underline{(c) The Need for BFs}: a drastic change in altitude may affect the stability of the UAV. 
\underline{(d) Example}: In this code snippet~\cite{citefunctiongvupdatereffromzdsp}
, the altitude change between two iterations of the control loop is bounded by \texttt{GV{\_}MAX{\_}Z{\_}DIFF}, the maximum distance between the previous waypoint and the current waypoint on the Z axis. \underline{(e) Occurrence}: 1 instance.

\pic{.5}{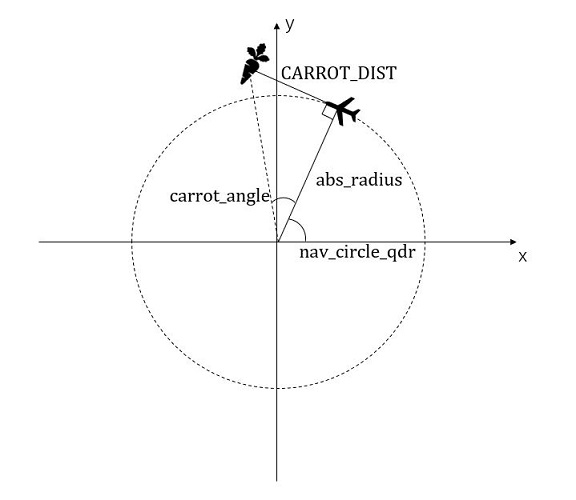}{Carrot-Based Guidance for Heading Change}{fig:carrotangle} 

\paragraph{Safe Heading Change in Guidance}
\label{subsection:headingmanagement}

\underline{(a) Use Context}: Paparazzi follows the widely used \emph{carrot-based} approach~\cite{citedynamic3dpath} for trajectory management: a virtual, continuously updated waypoint not far from the current position of the UAV to guide the next ``leg'' of movement of the UAV, similar to using a carrot to attract a mule to move forward. As shown in Figure \ref{fig:carrotangle}, the carrot-based guidance implemented by Paparazzi for circle navigation assumes a constant distance between the current position of the UAV and the carrot, as \texttt{CARROT{\_}DIST}. By adjusting the \texttt{carrot{\_}angle}, the UAV may change its heading. 
\underline{(b) Datatype}: heading (change).
\underline{(c) The Need for BFs}: a drastic change in heading may affect the stability of the UAV, and affects the correctness of the circle trajectory. 
\underline{(d) Example}: In Listing~\ref{code:navigation2c} which concerns circle navigation, the \texttt{carrot{\_}angle} is bounded to the range of $[\frac{\pi}{16}, \frac{\pi}{4}]$. The rest of the variables are illustrated in Figure \ref{fig:carrotangle}. \underline{(e) Occurrence}: 3 instances.

\begin{lstlisting}[language=C, caption={Safe Heading Change in Guidance~\cite{citefunctionnavcircle}}, captionpos=b, label={code:navigation2c}]
void nav_circle(struct EnuCoor_i *wp_center, int32_t radius)
{
  ...
      
  // direction of rotation
  int8_t sign_radius = radius > 0 ? 1 : -1;
  // absolute radius
  int32_t abs_radius = abs(radius);
  // carrot_angle
  int32_t carrot_angle = ((CARROT_DIST << INT32_ANGLE_FRAC) / abs_radius);
  Bound(carrot_angle, (INT32_ANGLE_PI / 16), INT32_ANGLE_PI_4);
  carrot_angle = nav_circle_qdr - sign_radius * carrot_angle;
  
  ...
}
\end{lstlisting}

\paragraph{Safe Leg Distance in Guidance}
\label{subsection:trajectorymanagement}

\underline{(a) Use Context}: For linear trajectories that do not involve heading change, Paparazzi also uses carrot-based guidance. In this setting, the distance between the starting point of the leg and the carrot, which is called \emph{leg distance}, is dynamically adjusted.  
\underline{(b) Datatype}: distance (X and Y axis).
\underline{(c) The Need for BFs}: if the leg distance is set too long, the UAV may go ``past'' the waypoint of the target point. Deviating from the planned trajectory is a correctness concern. 
\underline{(d) Example}: In this code snippet~\cite{citefunctionnavroute}
which concerns route (i.e., linear) navigation, the $\texttt{nav{\_}leg{\_}progress}$ is bounded to guarantee that the next leg of flight does not surpass the target waypoint. \underline{(e) Occurrence}: 4 instances.

\subsubsection{Sensor Management}
\label{section:sensormanagement}

As important components of a UAV, sensors play an irreplaceable role in UAV's state estimation, e.g., UAV's current attitude (pitch/roll/yaw). An accurate estimation based on sensor data is also critical for UAV safety. The physical variables related to sensor management are \emph{sensor readings}, the \emph{time interval} among readings, and the \emph{weight} when multiple sensor readings are weighted. We identified instances of BFs applied for sensor management, which we divide into 3 use scenarios.

\paragraph{Safe Sensor Readings}

\underline{(a) Use Context}: The raw sensing data may be unreliable, either because the sensor is faulty, or because the reading may only reflect a transient state.  
\underline{(b) Datatype}: sensor reading. 
\underline{(c) The Need for BFs}: The need for bounding is sensor-specific. Take the current sensor for example. Due to overflow on high current spikes (fast electrical transients in current), the reading may be magnitudes higher than normal readings. This would impact battery estimation, crucial for estimating the remaining flight time. 
\underline{(d) Example}: In this code snippet~\cite{citefunctionelectricalperiodic}
, the current sensor keeps its readings in \texttt{electrical.current}, 
which is in turn bounded to a safe range $[-65000, 65000]$. \underline{(e) Occurrence}: 1 instance.

\paragraph{Safe Sensor Reading Interval}

\underline{(a) Use Context}: In UAVs, sensors are continuously reading. In some scenarios, the time interval between different readings plays a crucial role in physical estimation. For example, as an application of Kalman filter~\cite{citekalman}, the UAV can use data from GPS and barometer at different time intervals to estimate its vertical position and velocity.
\underline{(b) Datatype}: time interval. 
\underline{(c) The Need for BFs}: if there is a significant delay between two intervals, the estimation may be inaccurate, which in turn severely impacts the decision-making process of the UAV.
\underline{(d) Example}: In this code snippet~\cite{citefunctioninsaltfloatupdategps}
, the variable \texttt{dt} represents the time interval between two GPS readings. It is bounded into the range $[0.02, 2]$ seconds. The variable is used by Kalman filter (\texttt{alt{\_}kalman}) for the estimation of the UAV's altitude and vertical speed.  \underline{(e) Occurrence}: 1 instance.

\paragraph{Safe Sensor Fusion}

\underline{(a) Use Context}: Complementary filter \cite{citecompfilter} combines sensor readings from the accelerometer and the gyroscope to estimate UAV attitude (pitch/roll/yaw).  
\underline{(b) Datatype}: weight for sensor fusion 
\underline{(c) The Need for BFs}: To ensure that data collected from both sensors are considered adequately, their proportions in attitude estimation need bounding in order to reach a balance between these two components. 
\underline{(d) Example}: In Listing \ref{code:ahrsfloatcmpl2c}, \texttt{ahrs{\_}fc.weight} computed at line 9 reflects the role of accelerometer plays in attitude estimation, which is influenced by \texttt{fabs(1.0 - g{\_}meas{\_}norm)}, the deviation between the measured gravitational acceleration and 1g. In the case of vibrations, large deviations from 1g may cause a decrement of the weight for the accelerometer data if bound is not introduced, ultimately causing the attitude estimate to drift \cite{citecompahrs}. 
Attitude estimation is critical for the safety of UAVs. 
In the aviation history, a catastrophe with the same root cause is Lion Air Flight 610, which was caused by incorrect angle-of-attack sensing (and consequent activation of the anti-stall software to repeatedly pitch the plane downward)~\cite{citelionair}.
\underline{(e) Occurrence}: 6 instances.

\begin{lstlisting}[language=C, caption= An Example of Sensor Fusion~\cite{citefunctionahrsfcupdateaccel}, captionpos=b, label={code:ahrsfloatcmpl2c}]
void ahrs_fc_update_accel(struct FloatVect3 *accel, float dt)
{
  ...
	
  // compute ratio between measured gravitational acceleration and the standard value
  const float g_meas_norm = float_vect3_norm(&filtered_gravity_measurement) / 9.81; 
  
  // compute the weight of accelerometer in attitude estimation
  ahrs_fc.weight = 1.0 - ahrs_fc.gravity_heuristic_factor * fabs(1.0 - g_meas_norm) / 10.0;

  Bound(ahrs_fc.weight, 0.15, 1.0);
  
  ...
}
\end{lstlisting}

\section{Understanding Bounding Functions Dynamically}
\label{section:evaluation}

In this section, we experimentally evaluate the impact of BFs on UAV behavior, answering \textbf{RQ2}. We start with a description of our rationale in \S~\ref{sec:rationdiff} and on experiment setup in \S~\ref{sec:setup}, and the core results from differential simulation will be described in the rest of the section with a summary and several more detailed case studies.

\subsection{The Rationale of Differential Simulation}
\label{sec:rationdiff}

As we stated earlier, UAVs are cyber-physical systems that interact with the physical world. 
In UAV software,  the traces of UAV physical properties --- pitch/roll/yaw, trajectory, or altitude as time series --- are essential for capturing their observable behavior. When the removal of BFs leads to observable difference in the trace of these physical variables, it should be a concern for attention.

Our differential simulation aims at achieving two goals. First, it helps confirm that the BF instances indeed impact the dynamic physical behavior of UAVs. A premise with the ``developer-in-the-field’’ approach is that we \emph{trust} the experience and wisdom of the developers. From the perspective, the dynamic approach here serves as the \emph{trust but verify} step: we would like to confirm BFs do make a difference in defining the physical behavior of UAVs. With that, answers to \textbf{RQ2} serve as an evidence of the significance of our taxonomy proposed for \textbf{RQ1}. Second, the dynamic approach also serves as a quantitative study of the safety-critical impact of BFs. It complements the qualitative study of our static (taxonomy) approach by answering \emph{how much} impact BFs have on the safety of UAV software.

\subsection{Experiment Setup} 
\label{sec:setup}

We use Paparazzi's built-in simulator for recording flight trajectories. We further use Paparazzi's log plotter to generate traces on real-time physical variables, such as speed, altitude, and roll-pitch-yaw values. The two complement each other, with the former useful for elucidating macro-level navigation patterns, and the latter useful for characterizing micro-level time-dependent physical behavior. 

Among the 109 BF instances, we are able to conduct simulation for 64 of them. Some programs with BF instances require manual radio control (RC) inputs. We have developed a script to ensure RC inputs are programmably given, so that for repetitions of the same experiment, identical RC commands with identical timing are inputed. The not-simulatable cases fall into two categories. First, the compilation and execution of some program fragments are hardware-dependent, such as requiring camera or sensor support. The Paparazzi simulator does support physical simulation, but it does not include features such as optical flow (for cameras) and some low-level sensors. Second, some code fragments where BFs occur are experimental features that cannot be built with any compatible aircraft. For example, no existing aircraft in Paparazzi is compatible with the module \texttt{stabilization\_float\_euler}, so we cannot simulate any BF instances in that module.

For each simulatable BF instance, we perform two experiments: (1) a simulation of the autopilot with a pre-defined flight plan (see \S~\ref{sec:navbk}) where the code with the BF instance is called; (2) a simulation with the same flight plan with the BF is removed. We compute whether the traces from the two experiments are different, where \emph{difference} is defined as the relative error in the trace values of physical variables (roll, pitch, yaw, thrust, etc.) from the two simulations above. We repeat each pair of simulations 5 times. The data across the 5 runs are averaged out with respect to timestamps.

It is noteworthy that when there is more than one BF instance in the same function, removing one may have no impact on the trajectory or physical variable traces, but removing multiple can. From now on, we refer to the experiments that involve the removal of multiple BFs in the same function at the same time as \emph{multi-BF differential simulation}, and refer to the one-BF-a-time experiments as \emph{single-BF differential simulation}. Multi-BF differential simulation is performed when single-BF differential simulation for each BF in a function does not show any difference.

\begin{table}
  \small
  \caption{Simulation Result Summary (S-Diff: Single-BF Simulation Different Results; M-Diff: Multi-BF Simulation Different Results; Same: No Difference in Results; Non-Sim: Not Simulatable)}
  \label{tab:simulationresults}
  \begin{tabular}{p{1.2cm}p{1cm}p{1cm}p{1cm}p{1.2cm}p{0.8cm}}
    \toprule
    Category & S-Diff & M-Diff & Same & Non-Sim & Total \\ 
    \midrule
    TM & 7 & 0 & 2 & 2 & 11 \\
    SM & 3 & 0 & 4 & 1 & 8 \\ 
    SAM & 5 & 4 & 8 & 0 & 17 \\
    EM & 2 & 0 & 5 & 14 & 21 \\
    PM & 3 & 6 & 15 & 28 & 52 \\
    \bottomrule
  \end{tabular}
\end{table}

\minipicfourbytwo{.19}{.01pt}{Relative Errors in Differential Analysis (Each sub-figure represents a distinct physical variable. Each bar represents a BF case, whose height is the mean and the range line is the standard deviation. The label to the left of each bar indicates a BF instance, whose details are described in Table~\ref{tab:diffcases}. Data is presented in log scale, where $10^0$ (1) implies 100\% relative error.)}{fig:relerrs}{cross_mean_std_x}{cross_mean_std_y}{cross_mean_std_alt}{cross_mean_std_course}{cross_mean_std_pitch}{cross_mean_std_roll}{cross_mean_std_speed}{cross_mean_std_climb}

\minipicfourbytwo{.18}{.01pt}{Pearson's Correlation Coefficients (PCCs) in Differential Analysis (Each sub-figure represents a distinct physical variable. Each bar represents a BF case, whose height is the PCC. The label at the bottom indicates a BF instance, whose details are described in Table~\ref{tab:diffcases}. A PCC value over 0.7 empirically indicates strong correlation.) }{fig:pccs}{cross_pcc_x}{cross_pcc_y}{cross_pcc_alt}{cross_pcc_course}{cross_pcc_pitch}{cross_pcc_roll}{cross_pcc_speed}{cross_pcc_climb}

\begin{table}
  \small
  \caption{Selected Differential Analysis BF Instances}
  \label{tab:diffcases}
  \centering
  \begin{tabular}{p{.7cm}p{3.3cm}p{3.5cm}}
    \toprule
    Label & File Name & BF Line Number \\ 
    \midrule
    A & {ahrs\_float\_cmpl} & 253\\
    B & {common\_nav} & 135\\
    C & {nav\_gls} & 150\\
    D & {nav\_gls} & 181\\
    E & {nav\_smooth} & 174\\
    F & {attitude\_ref\_saturate\_naive} & 79, 82, 83, 84 (multi-BF)\\
    G & {guidance\_h\_ref} & 240, 241 (multi-BF)\\
    \bottomrule
  \end{tabular}
\end{table}

\subsection{Result Summary}

The results from the experiments fall into 4 categories, which we summarize in Table~\ref{tab:simulationresults}. If our simulation shows difference in a single-BF differential simulation, we classify the involved instance as ``S-Diff''. Otherwise, if difference is shown in a multi-BF differential simulation, we classify the involved instances as ``M-Diff''. The rest of  simulatable instances are classified as ``Same'', and  non-simulatable instances are classified as ``Non-Sim''. There is no overlap between the categories. For repeated experiments, we only mark an instance as “different” when 5 repeated experiments all show a difference exists: in physical simulation, small variations are common, so we wish to be conservative to make sure all repeated experiments agree.

As we can see in the S-Diff and M-Diff columns, nearly half of the instances we can simulate produce different results when comparing executions with or without BFs. In other words, the BFs indeed play an important role in safeguarding the correctness of programs and consequently the safety of UAVs.

In our differential analysis, we compute the averaged relative error between the measured physical variable value of the program with the BF, and the one without. We elide yaw data for brevity. Figure~\ref{fig:relerrs} shows the result for a subset of BF instances, whose details can be found in Table~\ref{tab:diffcases}. The complete results are included in the repository.
Three concrete observations can be made. \emph{First}, BFs have non-equal impacts on physical variables. For example, we can observe that BF cases \texttt{A}, \texttt{B}, and \texttt{G} have large impacts on the majority of physical variables, whereas BF case \texttt{E} has a minor impact on nearly all variables. \emph{Second}, the same BF instance may have different impacts on different physical variables. For example, BF case \texttt{A} has a larger impact on the Z axis of positioning (altitude) (see Fig.~\ref{fig:relerrs-c}) than the Y axis (see Fig.~\ref{fig:relerrs-b}). As another example, the relative error of BF case \texttt{G} stands out in trajectory-related physical variables (see Fig.~\ref{fig:relerrs-d} for example) than speed-related variables. 
\emph{Third}, the same physical variable may be impacted by different BF instances in different degrees. For example, speed is significantly impacted by BF cases \texttt{A} and \texttt{B}, but not others (see Fig.~\ref{fig:relerrs-g}).

To gain a finer-grained analysis, we further computed the similarity of the two traces in a timestamp-wise manner.  
Figure~\ref{fig:pccs} shows the Pearson's Correlation Coefficients (PCCs) of the with-BF and without-BF traces. The most important observation is that PCC is rarely over 0.7, the golden standard for ``strong'' correlation. In other words, without BFs, a program would cause noticeable behavioral change to the UAV in a large number of BF instances, as manifested through location or pose or speed. 

Take BF case \texttt{E} for example. Recall that in the earlier relative error figure, this BF has a small relative error; the PCC results however tell a different story: 
a timestamp-wise alignment of traces is poor for the majority of physical variables, especially altitude, pitch, row, and speed on the X/Y dimensions. 
In this example, the BF is used to bound the physical variable of ground speed. With its removal, the UAV not only has significant ground speed fluctuation, but also leads to functuations in other physical variables. The difference between relative error and PCC as metrics is that the latter is time-dependent. As a result, PCC can capture behavior differences in the presence of (time-dependent) fluctuation despite the “mean” remains stable, a goal the relative error cannot achieve.

Together, these experiments show that BFs do significantly impact UAV behavior. As physical variables play a pivotal role in safety-critical UAV behavior -- from trajectory management to pose management and so on -- our experiments demonstrate the importance of BFs in safety-critical UAV software.

In the rest of this section, we highlight 3 BF instances and their impact on preserving the UAV behavior.

\subsection{Case Study: Turning Angles}

In \S~\ref{subsection:headingmanagement}, we discussed the bounded variable \texttt{carrot{\_}angle} bounded within the range $[\frac{\pi}{16}, \frac{\pi}{4}]$ in function \texttt{nav{\_}circle}, which is used by Paparazzi to perform a circle task. With the safeguard of the bounding function, the UAV circles around normally as is shown in Figure \ref{fig:turninganglecasestudy-a}, where the red actual trajectory fits nicely with the green desired trajectory.

However, if we remove the bounding function, as is shown in Figure \ref{fig:turninganglecasestudy-b}, the trajectory is irregular at the beginning and later follows a stable oval orbit. This deviation stems from the drastic angle variation \texttt{carrot{\_}angle}.

\minipic{.45}{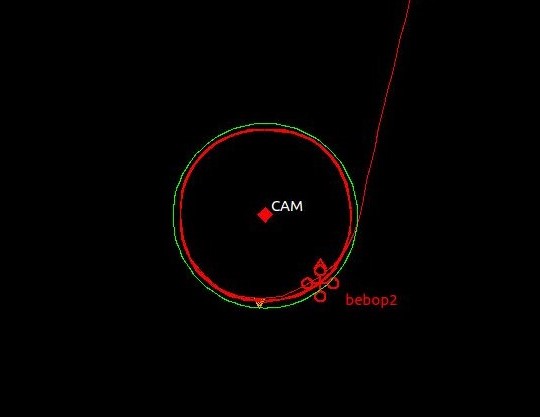}{5pt}{.45}{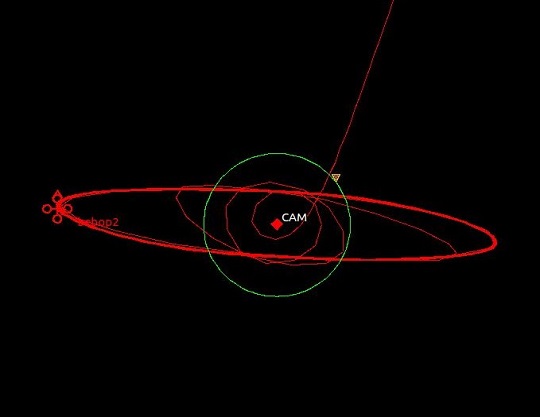}{A Case Study on Turning Angles (a) with-BF trajectory (b) without-BF trajectory}{fig:turninganglecasestudy}

\subsection{Case Study: Takeoff Speed}
\label{subsection:eval_takeoff_speed}

As an example of multi-BF differential simulation, Listing~\ref{code:energyctrlc3} shows a code snippet where removing only one BF does not make a difference while removing both does. In this example, \texttt{sp} represents the vertical speed setpoint computed by the PID algorithm, and \texttt{incr} represents its deviation from the current speed setpoint \texttt{v\_ctl\_climb\_setpoint}. \texttt{incr} is added to \texttt{v\_ctl\_climb\_setpoint} in the end. If we only remove the BF on line~\ref{line:bf_sp}, the excessive value would be bounded on line~\ref{line:bf_incr}. Similarly, if we remove the latter, since the former has already bounded \texttt{sp}, the following \texttt{incr} is thus not likely to be excessive. However, when we remove both, \texttt{v\_ctl\_climb\_setpoint} can grow by a sharp increment.

\begin{lstlisting}[language=C, caption={A Multi-BF Simulation Example~\cite{citefunctionvctlaltitudeloop}}, captionpos=b, label={code:energyctrlc3}, escapechar=|]
void v_ctl_altitude_loop(void)
{
  ...
  
  float sp = v_ctl_altitude_pgain * v_ctl_altitude_error + v_ctl_altitude_pre_climb ;

  BoundAbs(sp, v_ctl_max_climb); |\label{line:bf_sp}|

  float incr = sp - v_ctl_climb_setpoint;
  BoundAbs(incr, 2 * dt_navigation); |\label{line:bf_incr}|
  v_ctl_climb_setpoint += incr;
}
\end{lstlisting}

\pic{.75}{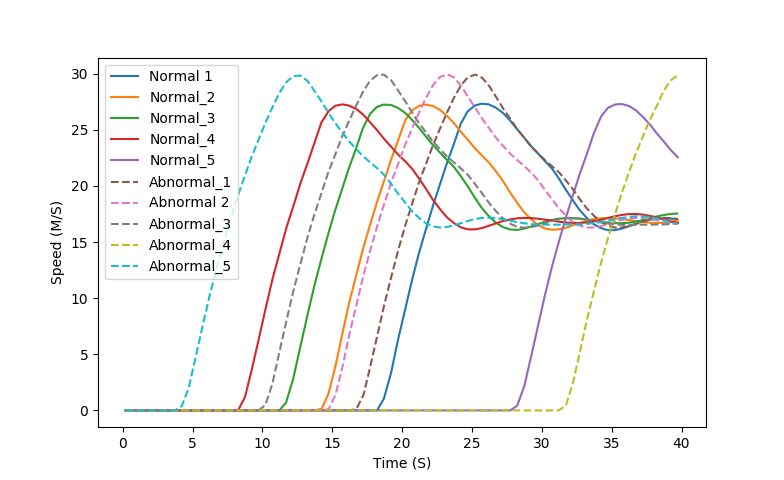}{Multi-BF Differential Simulation on Takeoff Speed}{fig:pairwisesimulationlog}

Figure~\ref{fig:pairwisesimulationlog} shows the UAV speed when taking off based on the UAV's flight logs. The solid lines show the speed when both BFs are kept, while the dashed line show the speed when both are removed. In the first 40 seconds, the without-BF runs (named as ``abnormal'' in the Figure) reach a higher speed during take-off: observe that the dashed lines show a higher speed than those of the solid lines. This agrees with our source code inspection above.

\subsection{Case Study: Navigation Progress}

In \S\ref{subsection:trajectorymanagement}, we discussed another bounded variable \texttt{nav{\_}leg{\_}progress} in function \texttt{nav{\_}route}, and this variable reflects the navigation progress which is bounded within the range $[0, \texttt{prog{\_}2}]$. As is shown in Figure \ref{fig:navprogcasestudy-a}, an oval trajectory consists of two straight lines and two semi-circles. The function \texttt{nav{\_}route} is called in the navigation task on both straight lines.  
If we remove the BF from variable \texttt{nav{\_}leg{\_}progress}, the UAV may ``flee'' and move in the opposite direction when approaching the waypoint where the straight routine begins, as is shown in Figure \ref{fig:navprogcasestudy-b}.

The more intriguing question is why the UAV would change its behavior as radically as this. Let us have a close look at the source code on how  \texttt{nav{\_}leg{\_}progress} is computed:

\begin{center}\texttt{nav\_leg\_progress = (pos\_diff.x * wp\_diff.x + pos\_diff.y * wp\_diff.y) / nav\_leg\_length;}\end{center}

\noindent Here, \texttt{wp{\_}diff} and \texttt{pos{\_}diff} are two-dimensional variables representing the horizontal distance between two waypoints \texttt{p1} and \texttt{p2} (as shown in the figure) and between the UAV and the start waypoint \texttt{p1} respectively. The types of their members \texttt{x} and \texttt{y} are both \emph{signed} integers. However, since \texttt{nav{\_}leg{\_}length} is an \emph{unsigned} integer, the result computed within the parentheses must be implicitly converted to unsigned integer before divided by \texttt{nav{\_}leg{\_}length}. When the UAV is approaching the waypoint \texttt{p1}, the result computed within the parentheses happens to be a negative value whose most significant bit is set to 1, and thus it is interpreted as a large value after conversion to the unsigned integer. After being divided by \texttt{nav{\_}leg{\_}length}, the most significant bit becomes 0 and therefore, when the final result is converted back to the signed integer and assigned to \texttt{nav{\_}leg{\_}progress}, it is still a large positive value which goes far beyond the range $[0, \texttt{prog{\_}2}]$. It further impacts the computation of the position of navigation target and consequently the UAV flees eccentrically. 

With a BF in place, \texttt{nav{\_}leg{\_}progress} is at least bounded within a range, so that a radically unexpected trajectory such as Figure \ref{fig:navprogcasestudy-b} does not occur. However, note that always adjusting its value to the upper bound \texttt{prog{\_}2} is not reasonable either: the variable should not have shown a completed progress before straight navigation begins. 

In other words, the program contains a bug. To help fix this bug, we added an explicit type conversion for \texttt{nav{\_}leg{\_}length}:

\begin{center}\texttt{nav\_leg\_progress = (pos\_diff.x * wp\_diff.x + pos\_diff.y * wp\_diff.y) / (int32\_t)nav\_leg\_length;}\end{center}

After this bug fix, we repeated our simulation without the BF. As is shown in Figure \ref{fig:navprogcasestudy-c}, the oval trajectory is preserved. However, as is analyzed in our discussion in \S~\ref{subsection:trajectorymanagement}, the carrot, namely the navigation target denoted as a yellow inverted triangle in the figure, exceeds the intended trajectory without the BF. 

This case study is interesting for two reasons. First, the use of the BF indeed reflects the developer's concern that an unbounded variable may significantly alter the UAV behavior. Second, the developer appears to be unaware of the latent bug: the BF use somewhat masks the severity of the bug. Observe however, it is the use of the BF that led our attention to this code snippet, and it is the simulation of BF removal that helps us uncover the bug. We reported this bug to Paparazzi, and our bug fix has been accepted. The updated code has now been merged into Paparazzi's GitHub repository.

\minipicthree{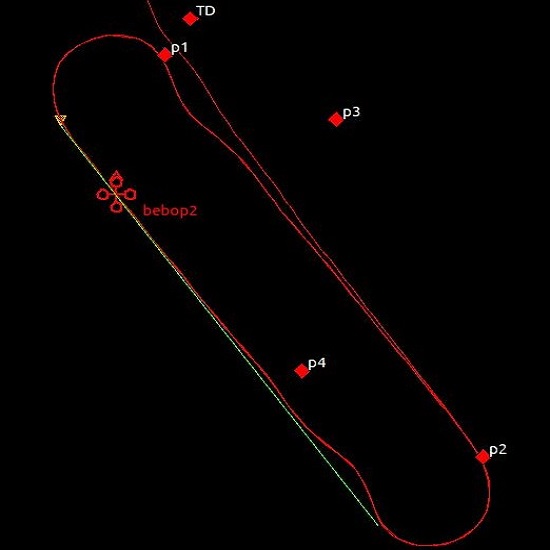}{oval_without_bound_fly_away.JPG}{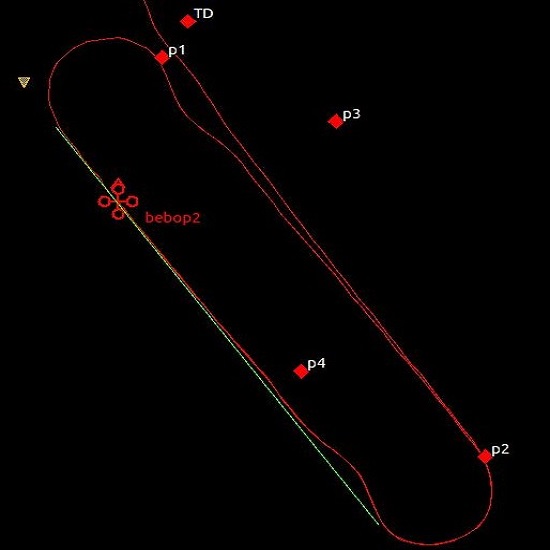}{.2}{10pt}{A Case Study on Navigation Progress (a) with-BF trajectory (original) (b) without-BF trajectory (original): UAV flees (c) without-BF trajectory (fixed): carrot off}{fig:navprogcasestudy}

\section{Threats to Validity}

Our analysis is empirical in nature.  
We based our analysis on the BFs injected by the Paparazzi developers. We assume the correct amount of functions is leveraged to achieve safety-criticality. In this sense, a fundamental limitation of our approach is that it may only be as good as the programming skills of the developers. In reality, developers may miss BFs and may make mistakes. Incompleteness in enumerating all safety-critical scenarios is inherent to our approach.

UAVs, and embedded systems in general, are real-time systems driven by their onboard sensors. As such, achieving simulations that faithfully cover all flight scenarios is challenging. Our simulation environment can replay navigation commands and replay at a specific rate.  However, due to timing of the control software, perfect reproducibility is impossible.

Paparazzi is an influential UAV framework, but not the only one. We believe the taxonomy of safety-critical use scenarios and the methodology of our differential analysis may transcend the specifics of Paparazzi, but the concrete findings of our study --- such as the number of use scenarios within each caregory, and the dynamic impact of individual cases --- may not be representative for all UAV frameworks.

\section{Related Work}
\label{section:relatedwork}

Verification for safety-critical software is a well-established area, and perhaps the best example of the ``top-down'' approach for studying safety of UAV systems. Blanchet et al.~\cite{blanchet-pldi03} propose a static analyzer based on abstract interpretation to verify a large class of properties in safety-critical software. Miller et al.~\cite{citenusmvforfcs} apply NuSMV\cite{citenusmv}, a symbolic model checker, to the verification of a flight control system. Kloetzer and Belta~\cite{kloetzer2008fully} provide a fully automated framework to develop feedback controllers for a linear system given its linear temporal logic over a set of linear predicates in its state variables. Kress-Gazit et al.~\cite{kress2009temporal} propose a linear temporal logic (LTL) based framework to automatically generate a hybrid controller that guarantees correct robot function given a high-level task specification as well as a class of admissible environments.
Yoo et al.\cite{citecasetool} introduce a formal-methods-based process that supports development, verification, and safety analysis for the nuclear power plant's reactor protection system, and develop Computer-Aided Software Engineering (CASE) tools for nuclear engineers to apply formal methods to safety verification.  Similarly, runtime verification of UAV software is also an actively researched topic. Moosbrugger et al.\cite{citer2u2} develop a real-time, Realizable, Responsible, Unobtrusive Unit (R2U2) to monitor security properties and diagnose security threats of Unmanned Aerial Systems (UAS) during run time. Its supervision scope covers the on-board components, as well as inputs from the ground control station.

Software engineering for self-adaptive cyber-physical systems is an active research direction, where UAVs are often cited as a compelling use scenario~\cite{Cheng2009,lemos13}. Testing cyber-physical systems (e.g.,~\cite{10.1145/3126521}) and development tools (e.g.,~\cite{10.1145/3183440.3190330,9283988}) is well explored. Another family of self-adaptive systems that have received attention in recent years is autonomous/self-driving vehicles, with results on bug characterization (e.g.,~\cite{DBLP:conf/icse/GarciaF0AXC20}) and testing (e.g.,~\cite{8453180,DBLP:conf/icse/ZhouLKGZ0Z020}). 
Programming languages are proposed for supporting energy awareness of UAVs~\cite{DBLP:conf/icse/LiuZ17} and context adaptation for UAVs~\cite{cop20}. Copilot~\cite{citecopilot} is a stream-based dataflow language to perform hard real-time monitoring over safety-critical control systems by sampling variables in programs and computing properties over the sampled values. SafetyScrum~\cite{citediscoveringsafetystories} is a software development methodology that relies on a notion of ``safety debt'' to incrementally track the safety status of safety-critical UAV systems in agile software development and maintenance.

Broadly speaking, our datatype-based classification can be related to programming language efforts that refine primitive types. For example, dimension types~\cite{Kennedy94dimensiontypes} are designed so that value 1 can either mean one meter or one kilometer, and misuse among them can be eliminated by the type system. As another example, Osprey~\cite{10.1145/1134285.1134323} is a constraint-based type inference to automatically detect misuse of measurement units. 

Fundamental to the growth of UAVs is their ability to fly autonomously and not require human control at all times. Most modern UAVs, from high-end fixed wing aircraft to hobby quadcopters, come equipped with flight controllers, such as in PixHawk~\cite{pixhawk}. 
These systems use well-studied algorithms such as extended Kalman filter estimation to fuse the sensor values into a pose, and well-studied controllers  to achieve the set commands. 
\section{Concluding Remarks}
\label{section:finaldiscussion}

This paper describes a novel empirical study on the use of bounding functions in UAV autopilot software. Our study shows that the use of bounding functions coincides with use scenarios where safety concerns of UAVs are addressed by UAV software developers. Our differential simulation further shows that bounding functions play an important role in preserving the physical behavior of UAVs. To the best of our knowledge, this is the first systematic empirical in-field study on open-source UAV software frameworks.

\textbf{Beneficiaries}
We envision our empirical study will be beneficial in the following ways.
(1) For \emph{UAV software developers}, our empirical study may serve as a reference point for systematically addressing safety concerns in future UAV development. UAVs are well known for their diverse hardware platforms, but the key safety-critical datatypes identified by this paper are likely to transcend the specifics of diverse platforms of UAVs. We show that despite the large code base, the BF instances revolve around a small set of physical variables, which future developers should pay particular attention to. 

(2) For \emph{framework and language designers}, our datatype-based taxonomy may inspire new abstractions to generalize, modularize, and reason about UAV software systems, with the identified datatypes and their associated use scenarios serving as motivations for new language-based designs such as automated BF placement and enforcement.

(3) For \emph{researchers} interested in \emph{automated analysis} for UAV software (e.g., through testing, debugging, and verification), our identified BFs and their differential simulation 
serve as a source for identifying new invariants, and as a litmus test on validating the coverage of their approaches. 
In addition, \texttt{PBF-Detector} can serve as a base system to facilitate Clang/LLVM-based development.

\textbf{Artifacts}
In the repository~\footnote{\label{modified_ppz_github}\url{https://github.com/SUNY-BU-Software-Systems-Research-Group/PaparazziBF}} 
, we provide the following artifacts: (a) the source code of \texttt{PBF-Detector} together with modified Paparazzi source (Makefiles); (b) a detailed documentation on each BF use; (c) all data of the simulation results, including log data, simulation screenshots, along with aircraft and flight plan files as test cases; (d) scripts for statistical analysis and for reproducing the results; (e) a report of the complete BF taxonomy.

\textbf{Acknowledgments}
We thank Brian Grant for his participation in the early stage of this project. We thank Gautier Hattenberger for his help on the Paparazzi Forum. This project is sponsored by NSF Awards CNS-1823260,
CNS-1823230, and SHF-1749539.

\bibliographystyle{IEEEtran}
\bibliography{citation.bib}

\begin{thebibliography}{10}
\providecommand{\url}[1]{#1}
\csname url@samestyle\endcsname
\providecommand{\newblock}{\relax}
\providecommand{\bibinfo}[2]{#2}
\providecommand{\BIBentrySTDinterwordspacing}{\spaceskip=0pt\relax}
\providecommand{\BIBentryALTinterwordstretchfactor}{4}
\providecommand{\BIBentryALTinterwordspacing}{\spaceskip=\fontdimen2\font plus
\BIBentryALTinterwordstretchfactor\fontdimen3\font minus
  \fontdimen4\font\relax}
\providecommand{\BIBforeignlanguage}[2]{{%
\expandafter\ifx\csname l@#1\endcsname\relax
\typeout{** WARNING: IEEEtran.bst: No hyphenation pattern has been}%
\typeout{** loaded for the language `#1'. Using the pattern for}%
\typeout{** the default language instead.}%
\else
\language=\csname l@#1\endcsname
\fi
#2}}
\providecommand{\BIBdecl}{\relax}
\BIBdecl

\bibitem{blanchet-pldi03}
\BIBentryALTinterwordspacing
B.~Blanchet, P.~Cousot, R.~Cousot, J.~Feret, L.~Mauborgne, A.~Min\'{e},
  D.~Monniaux, and X.~Rival, ``A static analyzer for large safety-critical
  software,'' \emph{SIGPLAN Not.}, vol.~38, no.~5, p. 196–207, May 2003.
  [Online]. Available: \url{https://doi.org/10.1145/780822.781153}
\BIBentrySTDinterwordspacing

\bibitem{citenusmvforfcs}
S.~Miller, E.~Anderson, L.~Wagner, M.~Whalen, and M.~Heimdahl, ``Formal
  verification of flight critical software,'' in \emph{AIAA Guidance,
  Navigation, and Control Conference and Exhibit}, 2005, p. 6431.

\bibitem{citenusmv}
A.~Cimatti, E.~Clarke, F.~Giunchiglia, and M.~Roveri, ``Nusmv: A new symbolic
  model verifier,'' in \emph{International conference on computer aided
  verification}.\hskip 1em plus 0.5em minus 0.4em\relax Springer, 1999, pp.
  495--499.

\bibitem{kloetzer2008fully}
M.~Kloetzer and C.~Belta, ``A fully automated framework for control of linear
  systems from temporal logic specifications,'' \emph{IEEE Transactions on
  Automatic Control}, vol.~53, no.~1, pp. 287--297, 2008.

\bibitem{kress2009temporal}
H.~Kress-Gazit, G.~E. Fainekos, and G.~J. Pappas, ``Temporal-logic-based
  reactive mission and motion planning,'' \emph{IEEE transactions on robotics},
  vol.~25, no.~6, pp. 1370--1381, 2009.

\bibitem{citecasetool}
J.~{Yoo}, E.~{Jee}, and S.~{Cha}, ``Formal modeling and verification of
  safety-critical software,'' \emph{IEEE Software}, vol.~26, no.~3, pp. 42--49,
  2009.

\bibitem{citer2u2}
P.~Moosbrugger, K.~Y. Rozier, and J.~Schumann, ``R2u2: Monitoring and diagnosis
  of security threats for unmanned aerial systems,'' \emph{Form. Methods Syst.
  Des.}, vol.~51, no.~1, p. 31–61, Aug. 2017.

\bibitem{r2u2-imav14}
G.~Hattenberger, M.~Bronz, and M.~Gorraz, ``Using the paparazzi uav system for
  scientific research,'' in \emph{{Proceedings of the International Micro Air
  Vehicle Conference and Competition 2014}}, August 2014.

\bibitem{citeadt}
B.~Liskov and S.~Zilles, ``Programming with abstract data types,'' \emph{ACM
  Sigplan Notices}, vol.~9, no.~4, pp. 50--59, 1974.

\bibitem{citefixedwingdefinition}
\BIBentryALTinterwordspacing
D.~Anderson and S.~Eberhardt. (2015) How airplanes fly: A physical description
  of lift. [Online; accessed 06-March-2020]. [Online]. Available:
  \url{http://www.aviation-history.com/theory/lift.htm}
\BIBentrySTDinterwordspacing

\bibitem{citerotarywingdefinition}
\BIBentryALTinterwordspacing
S.~May. (2017) What is a helicopter? [Online; accessed 06-March-2020].
  [Online]. Available:
  \url{https://www.nasa.gov/audience/forstudents/5-8/features/nasa-knows/what-is-a-helicopter-58.html}
\BIBentrySTDinterwordspacing

\bibitem{prw-overview}
\BIBentryALTinterwordspacing
{Pir Arkam}. (2020) How does a plane fly? [Online; accessed 9-May-2020].
  [Online]. Available:
  \url{https://rookieelectronics.com/the-aerodynamics-of-flight-how-does-a-plane-fly/}
\BIBentrySTDinterwordspacing

\bibitem{citepid}
\BIBentryALTinterwordspacing
M.~Araki, ``Pid control,'' in \emph{CONTROL SYSTEMS, ROBOTICS AND AUTOMATION -
  Volume II: System Analysis and Control: Classical Approaches-II},
  H.~Unbehauen, Ed.\hskip 1em plus 0.5em minus 0.4em\relax EOLSS Publications,
  2009, pp. 58--79. [Online]. Available:
  \url{https://books.google.com/books?id=RF1xDAAAQBAJ}
\BIBentrySTDinterwordspacing

\bibitem{citeindi}
E.~J.~J. Smeur, Q.~Chu, and G.~C. H.~E. de~Croon, ``Adaptive incremental
  nonlinear dynamic inversion for attitude control of micro air vehicles,''
  \emph{Journal of Guidance, Control, and Dynamics}, vol.~39, no.~3, pp.
  450--461, 2016.

\bibitem{aastrom1995pid}
K.~J. {\AA}str{\"o}m and T.~H{\"a}gglund, \emph{PID controllers: theory,
  design, and tuning}.\hskip 1em plus 0.5em minus 0.4em\relax Instrument
  society of America Research Triangle Park, NC, 1995, vol.~2.

\bibitem{citefunctionnavmovewaypoint}
``Function {nav\_move\_waypoint}: url at
  \url{https://github.com/paparazzi/paparazzi/blob/master/sw/airborne/subsystems/navigation/common_nav.c}.''

\bibitem{citefunctiongvupdatereffromzdsp}
``Function {gv\_update\_ref\_from\_zd\_sp}: url at
  \url{https://github.com/paparazzi/paparazzi/blob/master/sw/airborne/firmwares/rotorcraft/guidance/guidance_v_ref.c}.''

\bibitem{citedynamic3dpath}
G.~Conte, S.~Duranti, and T.~Merz, ``Dynamic 3d path following for an
  autonomous helicopter,'' in \emph{Proceedings of the 5th IFAC Symposium on
  Intelligent Autonomous Vehicles}, Oxford, UK, 2004, pp. 473--478.

\bibitem{citefunctionnavcircle}
``Function {nav\_circle}: url at
  \url{https://github.com/paparazzi/paparazzi/blob/master/sw/airborne/firmwares/rotorcraft/navigation.c}.''

\bibitem{citefunctionnavroute}
``Function {nav\_route}: url at
  \url{https://github.com/paparazzi/paparazzi/blob/master/sw/airborne/firmwares/rotorcraft/navigation.c}.''

\bibitem{citefunctionelectricalperiodic}
``Function {electrical\_periodic}: url at
  \url{https://github.com/paparazzi/paparazzi/blob/363dec86938cd1090221ccd772fc6fae58ed89a2/sw/airborne/subsystems/electrical.c}.''

\bibitem{citekalman}
R.~E. Kalman, ``A new approach to linear filtering and prediction problems,''
  1960.

\bibitem{citefunctioninsaltfloatupdategps}
``Function {ins\_alt\_float\_update\_gps}: url at
  \url{https://github.com/paparazzi/paparazzi/blob/master/sw/airborne/subsystems/ins/ins_alt_float.c}.''

\bibitem{citecompfilter}
W.~T. {Higgins}, ``A comparison of complementary and kalman filtering,''
  \emph{IEEE Transactions on Aerospace and Electronic Systems}, vol. AES-11,
  no.~3, pp. 321--325, May 1975.

\bibitem{citecompahrs}
\BIBentryALTinterwordspacing
{Paparazzi Wiki}. (2015) Vibration. [Online; accessed 10-February-2020].
  [Online]. Available:
  \url{https://wiki.paparazziuav.org/wiki/Vibration#Complementary_AHRS}
\BIBentrySTDinterwordspacing

\bibitem{citelionair}
\BIBentryALTinterwordspacing
D.~Shortell and J.~Shelley, ``Lion air crash investigators looking at two
  american companies associated with boeing 737 max sensor,'' \emph{CNN}, Apr
  2019. [Online]. Available:
  \url{https://www.cnn.com/2019/04/04/us/boeing-sensor-investigation/index.html}
\BIBentrySTDinterwordspacing

\bibitem{citefunctionahrsfcupdateaccel}
``Function {ahrs\_fc\_update\_accel}: url at
  \url{https://github.com/paparazzi/paparazzi/blob/master/sw/airborne/subsystems/ahrs/ahrs_float_cmpl.c}.''

\bibitem{citefunctionvctlaltitudeloop}
``Function {v\_ctl\_altitude\_loop}: url at
  \url{https://github.com/paparazzi/paparazzi/blob/master/sw/airborne/firmwares/fixedwing/guidance/energy_ctrl.c}.''

\bibitem{Cheng2009}
\BIBentryALTinterwordspacing
B.~H.~C. Cheng, R.~de~Lemos, H.~Giese, P.~Inverardi, J.~Magee, J.~Andersson,
  B.~Becker, N.~Bencomo, Y.~Brun, B.~Cukic, G.~Di~Marzo~Serugendo, S.~Dustdar,
  A.~Finkelstein, C.~Gacek, K.~Geihs, V.~Grassi, G.~Karsai, H.~M. Kienle,
  J.~Kramer, M.~Litoiu, S.~Malek, R.~Mirandola, H.~A. M{\"u}ller, S.~Park,
  M.~Shaw, M.~Tichy, M.~Tivoli, D.~Weyns, and J.~Whittle, \emph{Software
  Engineering for Self-Adaptive Systems: A Research Roadmap}.\hskip 1em plus
  0.5em minus 0.4em\relax Berlin, Heidelberg: Springer Berlin Heidelberg, 2009,
  pp. 1--26. [Online]. Available:
  \url{https://doi.org/10.1007/978-3-642-02161-9_1}
\BIBentrySTDinterwordspacing

\bibitem{lemos13}
R.~Lemos, H.~Giese, H.~Müller, J.~Andersson, M.~Litoiu, B.~Schmerl, G.~Tamura,
  N.~Villegas, T.~Vogel, D.~Weyns, L.~Baresi, B.~Becker, N.~Bencomo, Y.~Brun,
  B.~Cukic, R.~Desmarais, S.~Dustdar, G.~Engels, and J.~Wuttke, \emph{Software
  Engineering for Self-Adaptive Systems: A Second Research Roadmap}, 01 2013,
  pp. 1--32.

\bibitem{10.1145/3126521}
\BIBentryALTinterwordspacing
J.~Deshmukh, M.~Horvat, X.~Jin, R.~Majumdar, and V.~S. Prabhu, ``Testing
  cyber-physical systems through bayesian optimization,'' \emph{ACM Trans.
  Embed. Comput. Syst.}, vol.~16, no.~5s, Sep. 2017. [Online]. Available:
  \url{https://doi.org/10.1145/3126521}
\BIBentrySTDinterwordspacing

\bibitem{10.1145/3183440.3190330}
S.~A. Chowdhury, ``Automatically finding bugs in commercial cyber-physical
  system development tool chains,'' ser. ICSE '18, 2018, p. 506–508.

\bibitem{9283988}
S.~A. {Chowdhury}, S.~L. {Shrestha}, T.~T. {Johnson}, and C.~{Csallner},
  ``Slemi: Equivalence modulo input (emi) based mutation of cps models for
  finding compiler bugs in simulink,'' in \emph{2020 IEEE/ACM 42nd
  International Conference on Software Engineering (ICSE)}, 2020, pp. 335--346.

\bibitem{DBLP:conf/icse/GarciaF0AXC20}
J.~Garcia, Y.~Feng, J.~Shen, S.~Almanee, Y.~Xia, and Q.~A. Chen, ``A
  comprehensive study of autonomous vehicle bugs,'' in \emph{ICSE'20},
  G.~Rothermel and D.~Bae, Eds.\hskip 1em plus 0.5em minus 0.4em\relax {ACM},
  2020, pp. 385--396.

\bibitem{8453180}
R.~{Ben Abdessalem}, S.~{Nejati}, L.~{C. Briand}, and T.~{Stifter}, ``Testing
  vision-based control systems using learnable evolutionary algorithms,'' in
  \emph{2018 IEEE/ACM 40th International Conference on Software Engineering
  (ICSE)}, 2018, pp. 1016--1026.

\bibitem{DBLP:conf/icse/ZhouLKGZ0Z020}
H.~Zhou, W.~Li, Z.~Kong, J.~Guo, Y.~Zhang, B.~Yu, L.~Zhang, and C.~Liu,
  ``Deepbillboard: systematic physical-world testing of autonomous driving
  systems,'' in \emph{ICSE'20}, G.~Rothermel and D.~Bae, Eds.\hskip 1em plus
  0.5em minus 0.4em\relax {ACM}, 2020, pp. 347--358.

\bibitem{DBLP:conf/icse/LiuZ17}
Y.~D. Liu and L.~Ziarek, ``Toward energy-aware programming for unmanned aerial
  vehicles,'' in \emph{3rd {IEEE/ACM} International Workshop on Software
  Engineering for Smart Cyber-Physical Systems, SEsCPS@ICSE 2017, Buenos Aires,
  Argentina, May 21, 2017}.\hskip 1em plus 0.5em minus 0.4em\relax {IEEE},
  2017, pp. 30--33.

\bibitem{cop20}
J.~H. Burns, X.~Liang, and Y.~D. Liu, ``Adaptive variables for declarative uav
  planning,'' in \emph{The 12th International Workshop on Context-Oriented
  Programming and Advanced Modularity (COP)}, 2020.

\bibitem{citecopilot}
L.~Pike, A.~Goodloe, R.~Morisset, and S.~Niller, ``Copilot: a hard real-time
  runtime monitor,'' in \emph{International Conference on Runtime
  Verification}.\hskip 1em plus 0.5em minus 0.4em\relax Springer, 2010, pp.
  345--359.

\bibitem{citediscoveringsafetystories}
J.~{Cleland-Huang} and M.~{Vierhauser}, ``Discovering, analyzing, and managing
  safety stories in agile projects,'' in \emph{2018 IEEE 26th International
  Requirements Engineering Conference (RE)}, 2018, pp. 262--273.

\bibitem{Kennedy94dimensiontypes}
A.~Kennedy, ``Dimension types,'' in \emph{In 5th European Symp. on Programming,
  LNCS 788}.\hskip 1em plus 0.5em minus 0.4em\relax Springer-Verlag, 1994, pp.
  348--362.

\bibitem{10.1145/1134285.1134323}
\BIBentryALTinterwordspacing
L.~Jiang and Z.~Su, ``Osprey: A practical type system for validating
  dimensional unit correctness of c programs,'' in \emph{Proceedings of the
  28th International Conference on Software Engineering}, ser. ICSE
  ’06.\hskip 1em plus 0.5em minus 0.4em\relax New York, NY, USA: Association
  for Computing Machinery, 2006, p. 262–271. [Online]. Available:
  \url{https://doi.org/10.1145/1134285.1134323}
\BIBentrySTDinterwordspacing

\bibitem{pixhawk}
\BIBentryALTinterwordspacing
{PixHawk Team}. (2020) Pixhawk flight controller. [Online; accessed
  15-May-2020]. [Online]. Available: \url{https://pixhawk.org/}
\BIBentrySTDinterwordspacing

\end{thebibliography}

\end{document}